%% file: main_arXiv.tex
\documentclass[11pt,a4paper]{article}
\pdfoutput=1
\usepackage{jheppub}
\usepackage{ifpdf}
\usepackage[usenames,dvipsnames]{xcolor}
\usepackage{ae}
\usepackage[T1]{fontenc}
\usepackage[ansinew]{inputenc}
\usepackage{mathrsfs}
\usepackage{amsmath}
\usepackage{amssymb}
\usepackage{graphicx}
\usepackage{color}
\definecolor{darkblue}{cmyk}{0.9,0.9,0,0}
\definecolor{darkgreen}{rgb}{0,0.55,0}

\usepackage{epsfig}
\usepackage{amsmath}

\newcommand{\beq}{\begin{equation}}
\newcommand{\eeq}{\end{equation}}
\newcommand{\beqq}{\begin{equation*}}
\newcommand{\eeqq}{\end{equation*}}
\newcommand\beqa{\begin{eqnarray}}
\newcommand\eeqa{\end{eqnarray}}
\newcommand\beqaa{\begin{eqnarray*}}
\newcommand\eeqaa{\end{eqnarray*}}
\newcommand\bea{\begin{array}}
\newcommand\eea{\end{array}}

\def\XXint#1#2#3{{\setbox0=\hbox{$#1{#2#3}{\int}$ }
\vcenter{\hbox{$#2#3$ }}\kern-.5\wd0}}

\def\XXint#1#2#3{{\setbox0=\hbox{$#1{#2#3}{\int}$}
\vcenter{\hbox{$#2#3$}}\kern-.5\wd0}}

\newcommand{\nn}{\nonumber}

\newcommand{\neqa}{\nonumber\end{eqnarray}}
\newcommand{\la}[1]{\label{#1}}

\newcommand{\eq}[1]{(\ref{#1})}

\def\tr{{\rm tr~}}

\newcommand{\hs}{\frac{\sqrt{3}}{2}}
\renewcommand{\d}{\partial}

\newcommand{\<}{{\langle}}
\renewcommand{\>}{{\rangle}}

\newcommand{\cB}{{\cal B}}

\newcommand{\cL}{{\cal L}}

\newcommand{\re}{\relax{\rm I\kern-.18em R}}

\renewcommand{\sp}{p\hspace{-.40em}/}

\def\su2{{SU(2)}}

\def\[{\left[}
\def\]{\right]}

\def\s{\sigma}

\def\({\left(}
\def\){\right)}
\def\[{\left[}
\def\]{\right]}

\def\<{\langle}
\def\>{\rangle}

\def\cB{{\cal B}}

\def\cO{{\cal O}}

\def\s*{\ *_{\!\!\!\!\!\!\!\!\!\,_{\,_\text{\scriptsize{sym}}}}}
\def\hs*{\ \hat{*}_{\!\!\!\!\!\!\!\!\!\,_{\,_\text{\scriptsize{sym}}}}}
\def\d{\partial}

\def\i2{\frac{i}{2}}

\def\spi{\relax{\rm \pi\kern-0.5em /}}
\def\sA{\relax{\rm A\kern-0.5em /}}
\def\sp{\relax{\rm p\kern-0.5em /}}
\def\sd{\relax{\rm \d\kern-0.5em /}}
\def\sk{\relax{\rm k\kern-0.5em /}}
\def\sn{\relax{\rm n\kern-0.5em /}}
\def\sl{\relax{\rm l\kern-0.5em /}}
\def\sP{\relax{\rm P\kern-0.7em /}}
\def\sBethe{\relax{\rm \Bethe\kern-0.5em /}}
\def\cN{{\cal N}}

\numberwithin{equation}{section}

\usepackage{varioref}

\title{The Holographic Dual of Strongly $\gamma$-deformed N=4 SYM Theory: Derivation, Generalization, Integrability and Discrete Reparametrization Symmetry}
\author[a,b]{Nikolay Gromov}
\author[c,d]{Amit Sever}
\affiliation[a]{
Mathematics Department, King's College London,
The Strand, London WC2R 2LS, UK
}
\affiliation[b]{St.Petersburg INP, Gatchina, 188 300, St.Petersburg,
Russia}
\affiliation[c]{
School of Physics and Astronomy, Tel Aviv University, Ramat Aviv 69978, Israel}
\affiliation[d]{
CERN, Theoretical Physics Department, 1211 Geneva 23, Switzerland
}

\emailAdd{nikolay.gromov@kcl.ac.uk}
\emailAdd{amit.sever@cern.ch}
\abstract{Recently, we constructed the first-principle derivation of the holographic dual of ${\cal N}=4$ SYM in the double-scaled $\gamma$-deformed limit directly from the CFT side. The dual fishchain model
is a novel integrable chain of particles in $AdS_5$. It can be viewed as a discretized string and revives earlier string-bit approaches.
The original derivation was restricted to the operators built out of one of two types of scalar fields. In this paper, we extend our results to the general operators having any number of scalars of both types, except for a very special case when their numbers are equal. Interestingly, the extended model reveals a new discrete reparametrization symmetry of the ``world-sheet", preserving all integrals of motion. We use integrability to formulate a closed system of equations, which allows us to solve for the spectrum of the model in full generality,
and present non-perturbative numerical results. We show that our results are in agreement with the Asymptotic Bethe Ansatz of the fishnet model up to the wrapping order at weak coupling.
}
\begin{document}
\begin{flushright}
CERN-TH-2019-144
\end{flushright}
\maketitle
\newpage
\section{Introduction}

The strongly $\gamma$-deformed limit of $\cN=4$ SYM, also known as the {\it fishnet model}~\cite{Gurdogan:2015csr,Nielsen:1970bc,Zamolodchikov:1980mb}, is an ideal playground for developing our exact computational methods and understanding holography. It is an interacting four-dimensional QFT with $SU(N)$ symmetry that one could hope is solvable exactly in the planar limit. Just like the planar limit of ${\cal N}=4$ SYM theory, this model is conformal for any value of the 't Hooft coupling $\xi$. It consists of two complex $N\times N$ matrix scalars $\phi_1$ and $\phi_2$ that are weakly coupled when $\xi\to 0$. At strong coupling $\xi\to \infty$, a new classical dual description emerges in terms of a chain of scalars in $AdS_5$~\cite{Gromov:2019aku}. At finite $\xi$ the theory becomes rather complicated. At the same time, it is much simpler than ${\cal N}=4$ SYM. For example, at each order of the perturbation theory, there are only very few Feynman diagrams that contribute.

There are several properties that make the fishnet theory particularly interesting and tractable. First, the model can be proven to be integrable at the quantum level~\cite{Gurdogan:2015csr,Nielsen:1970bc,Zamolodchikov:1980mb}. Second, its Holographic dual fishchain theory can be derived rigorously and quantized exactly \cite{Gromov:2019aku,Gromov:2019bsj}. This contrasts with the current status of ${\cal N}=4$ theory, where the integrability is conjectured and the dual super-string in $AdS_5\times S^5$ can only be quantized semi-classically.

So far, our derivation and quantization of the fishchain model was restricted to a set of states/operators that belong to the so-called ${\mathfrak u}(1)$ sector. These are operators of the type $\cO=\tr[\phi_1^J\d^m(\phi_2\phi_2^\dagger)^n\dots]$, 
which can carry arbitrary spin and are only charged under the ${\mathfrak u}(1)_1$ subgroup of the total ${\mathfrak u}(1)_1\times {\mathfrak u}(1)_2$ symmetry that rotates the two types of scalars. The main objective of this paper is to extend the derivation to general states, carrying both quantum numbers. 
This involves introduction of the {\it magnons} i.e. $\phi_2$ fields in the background of $J$ $\phi_1$'s. While most of the steps go through very similar to our previous work, we find that the inclusion of the magnons leads naturally to a new notion of a {\it discrete reparametrization gauge symmetry}. 
It is the string-bit counterpart of the smooth string reparametrization freedom, which relates different ways of parametrizing the same state.\footnote{This symmetry is reminiscent of the Yangian symmetry of planar scattering amplitudes studied in \cite{Chicherin:2017frs}.} 
Furthermore, extending the integrability construction to all operators allows us, in particular, to solve the spectrum with magnons included, which is a new result by itself. The solution takes the form of a Baxter TQ-relations, subjected to the quantization conditions of \cite{Gromov:2019cja} and generalizing results of \cite{Gromov:2017cja}. 
It allows us to compute the conformal dimension of almost all single trace operators\footnote{There are some additional technical difficulties in deriving the spectrum for a special case when both $U(1)$ charges are equal.}. 
Moreover, this construction gives the general framework for development of the integraiblity based separation of variables method (SoV) for more general observables such as correlation functions. 

The paper is organized as follows. In section~\ref{sec2} we introduce our notations and review the main previous results, which are important for this paper. 
In particular, we recall the main steps from~\cite{Gromov:2019bsj} for the quantization procedure of the fishchain. In section~\ref{sec3}, we generalize the notion of the CFT wave function, 
initially introduced in~\cite{Gromov:2019bsj}, to operators with magnons. We then construct the corresponding graph building operator and derive its holographic counterpart. 
In section \ref{sec4}, we establish the integrability of the fishchain model with magnons. In section \ref{sec5}, we identify redundancy in the space of 
CFT wave functions that we associate to a discrete reparametrization symmetry of the fishchain model. In section \ref{sec:integ2}, we solve the spectrum of the model with magnons and prove its reparametrization symmetry. We conclude in the section \ref{sec7}.

\begin{figure}[t]
\centering
\includegraphics[scale=0.75]{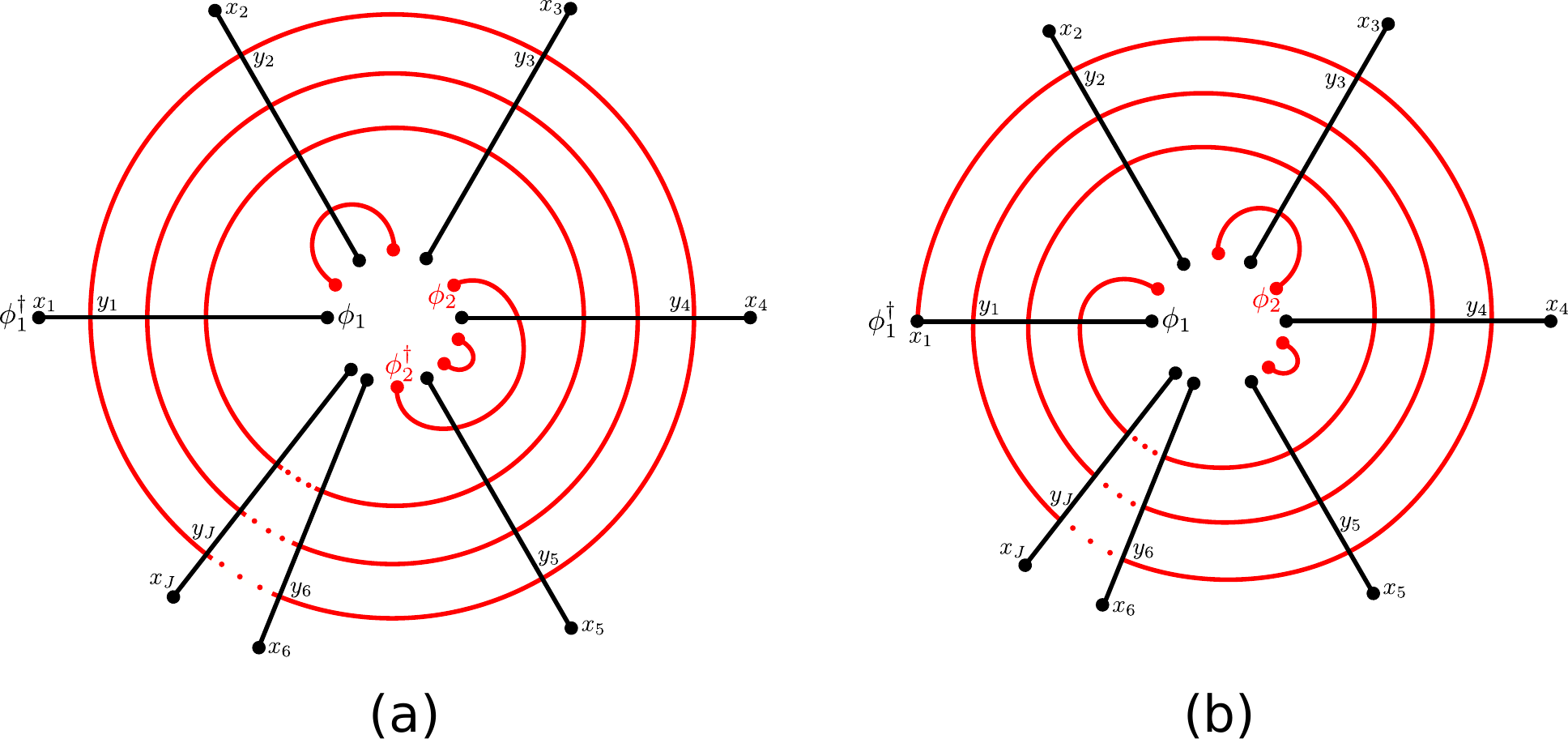}
\caption{a) Feynman diagrams that contribute to the correlation function of an operator of the form $\tr\!(\d^m\phi_1^J(\phi_2\phi_2^\dagger)^n\dots)$ are all of fishnet type -- made of the iterative wheel as shown on the diagram ~\cite{Nielsen:1970bc}. This structure can be resummed and leads to integrability \cite{Zamolodchikov:1980mb}. b) For operators with also net $\phi_2$ charge, (\ref{U2}), the diagrams are of iterative spider type~\cite{Caetano:2016ydc}.}
\label{fig:fishfig}
\end{figure}
\section{Notations and set-up}\la{sec2}
Here, we introduce the notations and 
give a short introduction to our results from \cite{Gromov:2019aku,Gromov:2019bsj}.
\subsection{Fishnet model}
The {\it fishnet model} \cite{Gurdogan:2015csr} is defined by the following action\footnote{We have suppressed double trace interactions which are not relevant non-perturbatively \cite{Fokken:2013aea,Sieg:2016vap,Gromov:2018hut}.}
\beq\la{fishnet}
\cL_{4d} = N\,{\rm tr}\(|\d\phi_1|^2+|\d\phi_2|^2
+ (4\pi)^2 \xi^2 \phi_1^\dagger\phi_2^\dagger\phi_1\phi_2\)\,.
\eeq
This model was obtained in \cite{Gurdogan:2015csr} by taking a double scaling limit of $\gamma$-deformed ${\cal N}=4$ SYM theory. Hence, we will still impose the Gauss constraint and will only consider operators that are neutral under the $SU(N)$ part of the global symmetries.

Our previous consideration was restricted to a subset of all operators, known as the ${\mathfrak u}(1)$ sector
\beq\la{U1}
\cO_J=\tr[\d^m\phi_1^J(\phi_2\phi_2^\dagger)^n\dots]+\dots
\eeq
containing any number of derivatives, $J$-scalar fields $\phi_1$ and any {\it neutral} combination of $\phi_2$ and $\phi_2^\dagger$. Correlation functions of ${\mathfrak u}(1)$ operators are given by the sum of fishnet type diagrams. These are diagrams that are made of the iterative wheel structure as is demonstrated in figure \ref{fig:fishfig}.a.

General single trace operators in the model can also carry ${\mathfrak u}(1)_2$ charge. They take the schematic form
\beq\la{U2}
\cO_{J_1}=\tr[\d^{m_1}\phi_1^{J_1}\d^{m_2}\phi_2^{J_2}(\phi_2\phi_2^\dagger)^{n_1}(\phi_1^\dagger\phi_1)^{n_2}\dots]+\dots\;.
\eeq
In this paper we allow for arbitrary values of $J_1$ and $J_2$ as long as $|J_1|\neq |J_2|$. Surprisingly, this case seems to be very different as one can see already in the explicit results of~\cite{Gromov:2018hut}.

Correlation functions of these type of operators consist of a very specific type of planar Feynman diagrams that have the shape of a spider web. In figure \ref{fig:fishfig}.b we draw an example of such a diagram for the case where $J_2=1$. In general, the diagrams can be constructed by first contracting all the $\phi_1$ fields in the trace, (the black lines in figure \ref{fig:fishfig}). Then, the $\phi_2$'s have no other option but to spiral around the trace, crossing the $\phi_1$ lines from the right to the left, (the red line in figure \ref{fig:fishfig}.b). Each time a $\phi_2$ line crosses a $\phi_1$ line we have the four scalars interaction of the fishnet model (\ref{fishnet}). The absence of the complex conjugate interaction vertex forbids any other diagram to contribute in the planar limit.

\subsection{Fishchain model}\la{secFC}
\begin{figure}[t]
\centering
\includegraphics[scale=0.9]{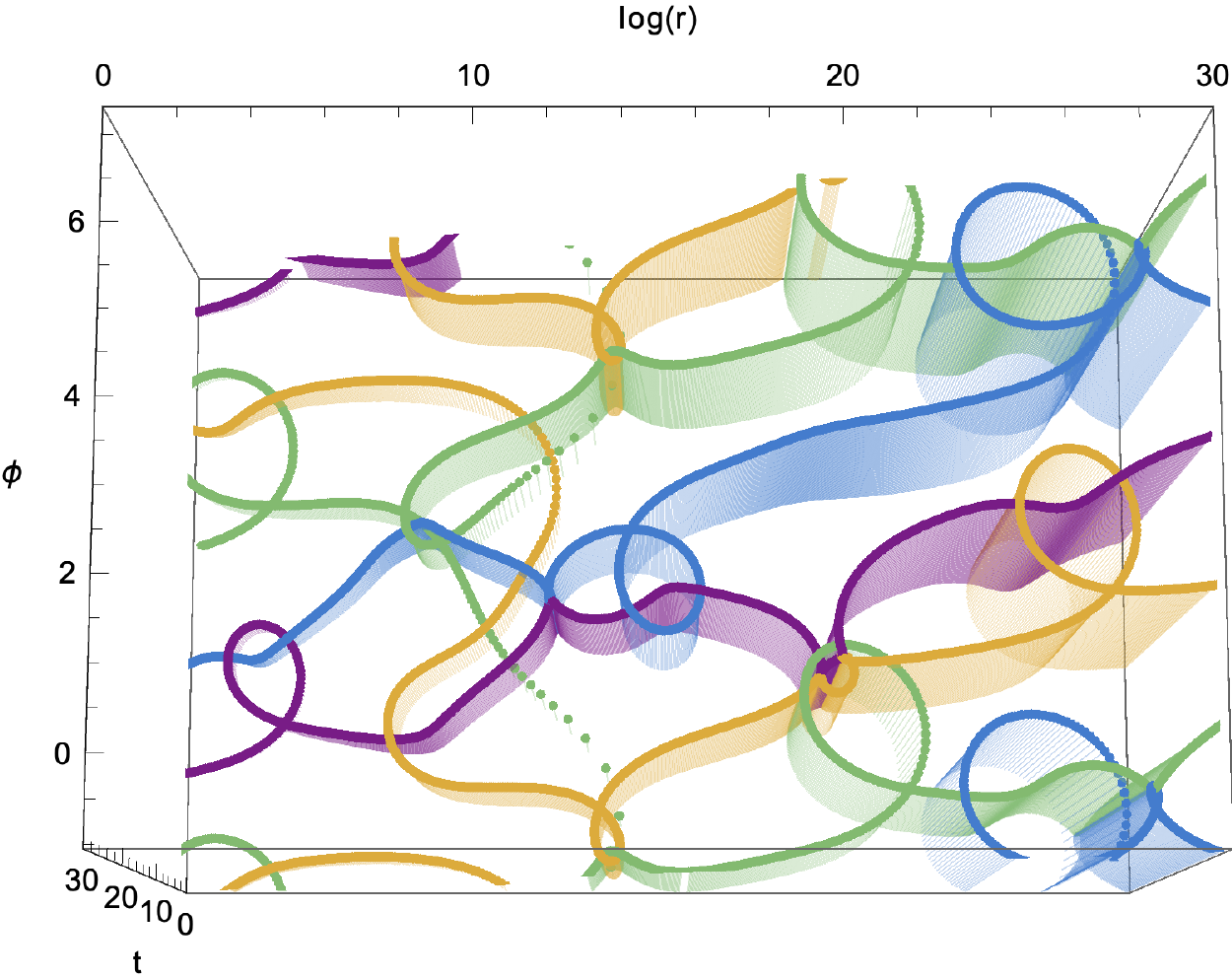}
\caption{Dynamics of the fishchain consisting of $J=4$ particles (denoted by different colours). The initial conditions are restricted to a $2D$ plane after projection on the boundary. The plane is parametrized in the radial parameters by the angle $\phi$ and the radius $r$. The third axis corresponds to the time. The drift in $\phi$ is due to a non-zero total angular momentum $S$ and the exponential expansion in $r$ is due to $\Delta>0$.}
\label{fig:particles}
\end{figure}
The holographic fishchain dual of the ${\mathfrak u}(1)$ operators of charge $J_1$ consists of a chain of $J_1$ point-like scalar particles in $AdS_5$. In the classical/strong coupling limit, $\xi\to\infty$, the theory is defined by the action
\beq\la{Sdual}
S_{\rm dual}=\xi \int dt \sum_{i=1}^{J_1}\[
\frac{\dot X_i^2}{2}+\prod_k\(-X_k.X_{k+1}\)^{-\frac{1}{J_1}}
\]\;,
\eeq
where $X_i\in{\mathbb R}^{1,5}$. While at the quantum level the particles live on $AdS_5$ space, at the classical limit its radius  collapses to zero and the dynamics take place on the light cone, $X_i^2=0$. One can think about \eq{Sdual} as a discretized version of the gauge fixed Polyakov action for a string, which requires a Virasoro condition.
An analog of the Virasoro condition in the case of the fishchain is $\dot X_i^2=1$. 
In addition to the constraints
\beq\la{constraints}
X_i^2=0\ ,\qquad\dot X_i^2=1\ ,\qquad i=1,2,\dots,J_1\;
\eeq
the Hamiltonian of the system should vanish to ensure the time reparametrization symmetry
\beq\la{constraints2}
H_q=0\ ,\qquad H_q=\tr \(\frac{q_J^2}{2}\dots
\frac{q_1^2}{2}\)-1\;,
\eeq
where $q_i$ is the local charge density $q_i^{MN}\equiv 
\dot X_i^M X_i^N-\dot X_i^N X_i^M$.
The classical Hamiltonian $H_q$ follows from the action \eq{Sdual} together with the constraints (\ref{constraints}), see \cite{Gromov:2019bsj}.

As can be seen from figure \ref{fig:particles}, the action \eq{Sdual} produces rather non-trivial dynamics. Even though this dynamic may look chaotic, the model is integrable.

\paragraph{Quantization.}
We now briefly go through the main steps of the canonical quantization procedure. 
We will follow the Dirac procedure for the quantization of a system with constraints.
To begin with, we notice that in addition to the primary constraints \eq{constraints}
we also have to include the secondary constraint
$\dot X_i. X_i=0$. Using that the canonically conjugate
to $X_i$ variable is $P_i=\dot X_i$ we have the following constraints
\beq\la{constraintsPX}
\varphi_{i,X}=X_i^2\ ,\qquad \varphi_{i,P}=P_i^2-1\ ,\qquad \varphi_{i}=X_i.P_i\qquad\text{and}\qquad  H_q\ .
\eeq
Among these, $\varphi_j$ and $\varphi_{j,P}$ are second class constraints as follows from their Poisson bracket. Hence, in the quantization of the model we first introduce corresponding Dirac brackets and only then promote them into quantum commutation relations. For example, the commutator between the operators $\hat X$ and $\hat P$ takes a non-standard form
\beq
[\hat X_k^M,\hat P_j^N]=
\frac{i\delta_{kj}}{\xi}\(\eta^{MN}-\hat P_k^M \hat P_K^N\)\;.
\eeq
Similarly, the commutator between different components of $\hat X_k^M$ is no longer zero. To resolve these commutation relations we introduce the operator $\hat Y_i^N=\frac{i}{\xi}\d_{P_{i,N}}$, which commutes in the standard way with $\hat P_{k}^N$. Then we show that we automatically solve all commutation relations
if 
$\hat X_i^M=\hat Y_i^M-(\hat Y_i.\hat P_i)\;\hat P_i^M$. Next, we have to impose the constraints~\eq{constraintsPX} at the quantum level and resolve possible ordering ambiguities.
What we find is that 
\begin{itemize}
    \item The quantum constraint $\hat\varphi_i=\hat X_i.\hat P_i$ is automatically satisfied. Here, we had to choose an ordering between $\hat X_i$ and $\hat P_i$ in $\hat\varphi_i$. To satisfy the constraint, we had to correlate this choice with the one in the definition of $\hat X_i$ in terms of $\hat Y_i$.
    \item The constraint $\hat\varphi_{i,X}$ is proportional to $\tr q_i^2$, which is nothing but the quadratic Casimir, $\hat{\bf C}_{i,2}/\xi^2$. Instead of setting it to zero we identify it with the dimension $1$ of the scalar field $\phi_1$, corresponding to $\hat{\bf C}_{i,2}=-3$.
    \item The constraint $\hat\varphi_{i,P}$ fixed the dependence of the wave function on $R_i^2\equiv -Y_i.Y_i$, as a result all the dynamical information is contained in the reduced wave-function $\Psi(Z_1,\dots,Z_J)$ depending on $J$ coordinates $Z_i^M$ on $AdS_5$ that are related to the $Y_i$'s via $Y_i^M=R_i Z_i^M$.
    \item Finally, we have to resolve the ordering ambiguity in the quantum version of $H_q$.
    Since the matrix elements of $\hat q_i^{NM}$ do not commute with each other, 
    the quantum version of $q_i^2$ suffers from ambiguities. As we explain below, the relevant definition is the symmetric and traceless combination 
    \beq\la{qno}
    :\hat q_i^2:\,=\hat  q_i^2-\frac{2i}{\xi}\hat  q_i-\frac{1}{\xi^2}\;,
    \eeq
    where the charge density in the $Z$-coordinates is  
    \beq\la{qZ}
    \hat q^{MN}_j=-{\frac{i}{\xi}}(Z_j^N\d_{ Z_{j\,M}}-Z_j^M{\frac{\d}{\d Z_{j\,N}}})\ .
    \eeq
    The corresponding quantum fishchain Hamiltonian takes the form
    
\beq\la{Handnq}
\hat H_q = 
\tr\(  \frac{:\hat{ q}_J^2:}{2}\dots \frac{:\hat{ q}_1^2:}{2}\)-1\;.
\eeq
\end{itemize}

At the end of the day we end with a quantum theory in $AdS_5$ that is defined by two relations that the fishchain wave function $\Psi(Z_1,\dots,Z_J)$ has to satisfy
\beq\la{Hqq}
\hat H_q\circ \Psi=0\ ,\qquad\text{and}\qquad\tr \hat q_i^2
\circ \Psi=\frac{2\Delta_i(4-\Delta_i)}{\xi^2} \Psi
\eeq
where $\Delta_i=1$ is the dimension of the scalar $\phi_1$.

\paragraph{The Holographic map.}
The key object which allows to establish the Holographic duality between the fishchain model and the fishnet CFT is the {\it CFT wave function} introduced in \cite{Gromov:2019bsj}. This is a $J+1$ point correlator of a local operator $\cO$ and $J$ scalar fields
\beq\la{wf0}
\varphi_\cO(x_1,\dots,x_J)=\<\cO(x_0)\,\tr[\phi^\dagger_1(x_1)\dots\phi^\dagger_{J}(x_J)]\>\ .
\eeq
The knowledge of this function is equivalent to specifing the local operator $\cO$. In order to compute $\varphi_\cO$ in perturbation theory one has to sum an infinite number of wheel diagrams like in figure~\ref{fig:fishfig}(a). To sum these diagrams one can use the so-called graph building operator $\hat \cB$, which is an operator acting on the wave function $\varphi_\cO$ by adding one more wheel to the diagrams
\beq\la{Bhat}
\hat {\cal B}\circ f(\vec x_1,\dots,\vec x_J)=\int
\prod_{i=1}^Jd^4 y_i \prod_{j=1}^J
\frac{\xi^2/\pi^2}{(\vec y_i-\vec y_{i+1})^2(\vec y_j-\vec x_j)^2}f (\vec y_1,\dots,\vec y_J)\;.
\eeq
Physical non-protected operators in the CFT spectrum are in one to one correspondence with stationary wave functions 
\beq\la{Bphi}
\hat \cB\circ \varphi_\cO(x_1,\dots,x_J)=
\varphi_\cO(x_1,\dots,x_J)\;.
\eeq
The map between the CFT and the fishchain wave functions reads \cite{Gromov:2019bsj} 
\beq\la{bwf}
\psi_{\cal O}^{\bf I}(Z_1,\dots,Z_J)=
\int \prod_{i=1}^J \frac{D^4 X_i}{-4\pi^2(Z_i.X_i)^{3}}
\Phi_\cO(X_1,\dots,X_J)\;,
\eeq
where $X^{M=-1,\dots,4}$ are six dimensional embedding coordinates with $(-,+,+,+,+,+)$ signature \cite{Dirac:1936fq}. They realize flat four dimensional space as the protective lightcone of ${\mathbb R}^{1,5}$ as
\beq\la{embeddingX}
X={\frac{1}{2}}X^+\(1+x^2,1-x^2,2\vec x\)\;.
\eeq
In this parameterization we have for example $1/(x-y)^2=X^+Y^+/(-2X.Y)$. When rewriting the correlation functions in embedding space we are following the general prescription of \cite{SimmonsDuffin:2012uy}. According to it we have to introduce extra factors of $X^+$ in agreement with the scaling dimensions of the corresponding operators. In particular, the embedding space wave function in (\ref{bwf}) is defined as 
\beq\la{embeddingwf1}
\Phi_\cO(X_1,\dots,X_J)\equiv{
\frac{
\varphi_\cO
(\vec X_1/X_1^+,\dots,\vec X_J/X_J^+)}{ X_1^+\,\dots\,X_J^+\,(X_0^+)^{\Delta_\cO}}}\;.
\eeq
The introduction of these extra factors ensures that the r.h.s. stays invariant under all conformal transformations.

It was shown in \cite{Gromov:2019bsj} that if $\phi_\cO$ obeys the condition \eq{Bphi}, then the l.h.s. of \eq{bwf} will be automatically annihilated by $\hat H_q$. At the same time the second condition in \eq{Hqq} is also satisfied just because the kernel $\frac{1}{(Z.X)^3}$ solves this equation for any null vector $X$. In the next section we will show how this Holographic map naturally generalizes to the case where both quantum numbers $J_1=J$ and $|J_2|<J$ are non-zero.

\section{CFT wave function and the graph building operator}\la{sec3}

In this section we generalize the fishchain model to include magnons i.e. insertions of the $\phi_2$ fields into the ``vacuum" of $\phi_1$'s. We assume that the number of $\phi_2$ fields is less the number of $\phi_1$'s. The starting point for the construction is the {\it CFT wave function}, which we  introduced in \eq{wf0} for the $\mathfrak{u}(1)$ sector. Like before it is given by the $(J+1)$-point correlation function between a single trace operator of the type (\ref{U2}) and the trace of $J$ local operators, see figure \ref{fig:fishfig}.b
\beq\la{wf}
\varphi^{\bf I}_\cO(x_1,\dots,x_J)=\<\cO(x_0)\,\tr[\chi_{I_1}(x_1)\dots\chi_{I_J}(x_J)]\>\ .
\eeq
The new ingredient is that 
the local operators $\chi_a$ are not just
single scalars $\phi_1^\dagger$, but come in two types
\beq\la{chis}
{\includegraphics[scale=0.5]{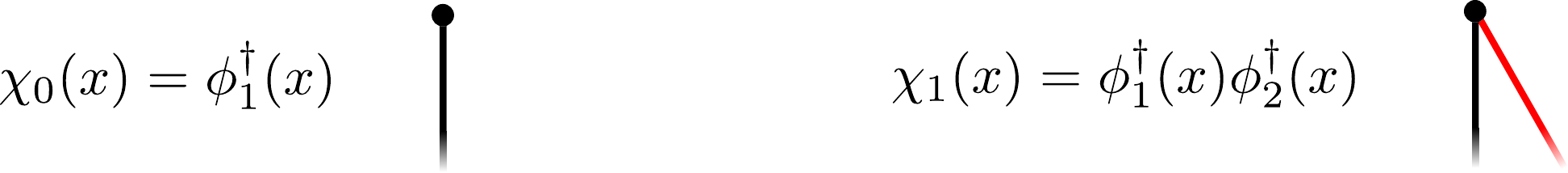}}\ .
\eeq
We use the multi-index notation ${\bf I}=(I_1,\dots,I_J)$ where the index  $I_j\in\{0,1\}$.
For example, when $I_j=1$ it indicates the position of the $\chi_1$ insertion, which is a composite of two scalars $\phi_1^\dagger$ and $\phi_2^\dagger$. 
Like before, the knowledge of this correlator is equivalent to the knowledge of the initial operator. Any other planar correlators, such as those with $\phi_2\phi_2^\dagger$ and $\phi_1^\dagger\phi_1$, can be expressed in terms of this one. 
In section \ref{sec5} we show that the order of the magnons inside the correlator can be interchanged. Hence, in order to reconstruct the local operator it is sufficient to know the CFT wave function for only one of all possible magnon orderings. In this section we consider all magnon orderings on equal footing.

As for the case with no magnons, the Feynman diagrams which contribute to the correlator $\varphi^{\bf I}_\cO$ in (\ref{wf}) are of iterative structure and can be represented as a simple geometric series built from the corresponding graph building operator $1/(1-\hat\cB)$. 
The only difference in comparison to the previous section is that now the graph building operator $\hat\cB$ is slightly more complicated and is given by
\beq\la{graphbuilding}
\hat\cB\circ f(x_1,\dots,x_J)\equiv \int\prod_{i=1}^Jd^4y_i\(\prod_{j=1}^Jb_{I_j}(x_j,y_j,y_{j-1})\)\,
f(\vec y_1,\dots,\vec y_J)\ ,
\eeq
\begin{figure}[t]
\centering
\def\svgwidth{11cm}

\ifpdf
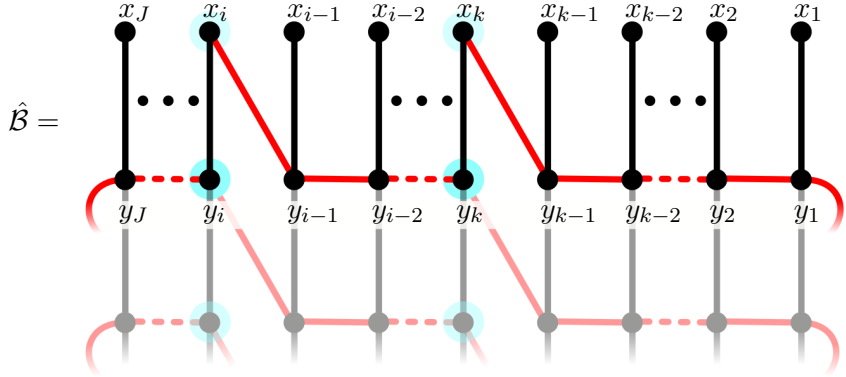
\else
\com{PDF-picture replacement}
\fi

\caption{\small The graph-building operator $\hat{\cB}$ for the state with two magnons. The locations of the magnons are indicated by the blue circles.}\label{fig:graphs_magnons}
\end{figure}
where $f$ is a probe function. In \eq{fig:graphs_magnons} the integration kernel is built out of elementary building blocks $b_{a}$,
which are different for the sites with magnons and without
\beq\la{kernel}
b_I(x,y,\tilde y)={\frac{\xi^2}{\pi^2(x-y)^2}}\times\left\{\begin{split}
{1\over(y-\tilde y)^2}&\qquad I=0\;\;\text{(no magnon)}\\
{1\over(x-\tilde y)^2}&\qquad I=1\;\;\text{(magnon)}\end{split}\right.\;.
\eeq
We see that at the sites with a magnon, the $\phi_2$ propagator connects to the external point $x_j$ instead of the integrated $y_j$ point as in a site without a magnon. 
This is illustrated in  figure \ref{fig:graphs_magnons}, where the sites with magnons are highlighted by the blue circles.

As we explained in the previous section 
the key equation which the CFT wave function should satisfy
is \eq{Bphi}.
The wave function is the correlator computed at some finite value of the coupling $\xi$. It is given by an infinite sum of planar Feynman diagrams. As such it should stay invariant under addition of an extra layer i.e. the wave function should be an eigenstate of the graph building operator with a unit eigenvalue, (\ref{Bphi}). 
For explicit examples we refer to \cite{Gromov:2018hut}. 
As was demonstrated in \cite{Gromov:2018hut}
the condition \eq{Bphi} singles out the physical 
states in the theory and 
leads to the discrete spectrum of the scaling dimensions $\Delta$ corresponding to the properly regularized and renormalized single trace operators of the type (\ref{U2}). 
Like in our previous paper \cite{Gromov:2019bsj}, we will now rewrite \eq{Bphi} in a form suitable for uplifting to $AdS_5$ space and show that the dynamics induced by \eq{Bphi} is exactly this of the quantum fishchain (QFC).

\subsection{Inverting the graph building operator}
Following \cite{Gromov:2019bsj}
in order to relate the graph building operator $\cal B$ to the dual fishchain model in $AdS_5$, we have to invert it and rewrite \eq{Bphi} as a zero energy Hamiltonian constraint
\beq\la{phys}
\hat H\,|\varphi_{\cal O}\rangle\equiv\(\hat\cB^{-1}-1\)|\varphi_{\cal O}\rangle=0\;.
\eeq
This constraint is then interpreted as the result of a time reparametrization symmetry of the fishchain. 
We show below that, defined in this way, $\hat\cB^{-1}$ and $\hat H$ are differential operators. 

In the case without magnons the invertion of $\hat\cB$, given by \eq{Bhat}, was straightforward. We simply act with $\Box_j$ on all the external $x_k$ points in (\ref{Bhat}). This has the effect of eliminating the corresponding $y_k$ integrals and the $1/(4\pi^2(x_i-y_i)^2)$ factors.\footnote{Recall that $\Box_i\frac{1}{4\pi^2(x_i-y_i)^2}=-\delta^4(x_i-y_i)$.} In the case at hand we have an extra complication -- there are two propagators that end at a site with a magnon.
Acting with $\Box_j$ on the magnon site does not remove the integration anymore,  see figure \ref{fig:graphs_magnons}.

\begin{figure}[t]
\centering
\def\svgwidth{15cm}

\ifpdf
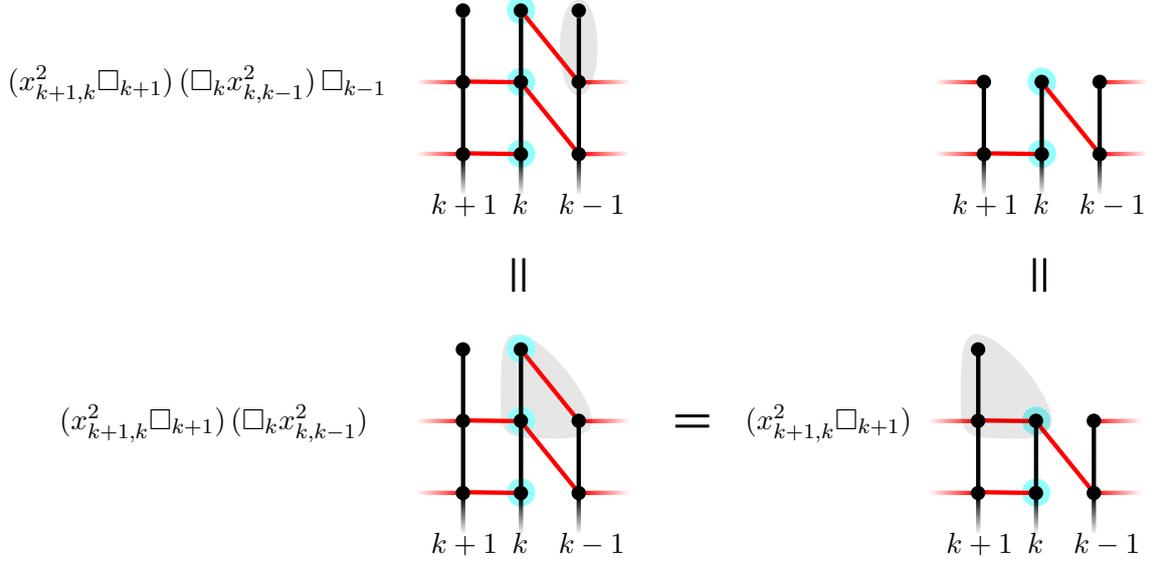
\else
\com{PDF-picture replacement}
\fi

\caption{In the presence of a magnons, the inverse of the graph building operator is more involved. Here, we demonstrate how the operator is inverted for the case where there is a magnon at the $k$'th site but no magnons at the neighboring sites.}
\label{magnon_move2}
\end{figure}

Let us show that the inversion of ${\cal B}$ is still possible in the situation when there is at least one site without a magnon, i.e. for $J_2<J_1$. Without loss of generality, we assume that there is a magnon on the $k$'th site, but there is no magnon at the neighboring site to the right i.e. $I_{k-1}=0,\; I_k=1$, see figure \ref{magnon_move2}. In this case we can hit $x_{k-1}$ with $\Box_{k-1}$, removing the integration in $y_{k-1}$. After that
the propagator connecting $y_{k-1}$ (which is now equal to $x_k$) with $x_k$ can be removed by multiplication with $x_{k,k-1}^2$.\footnote{We use the standard notation $x_{ab}=x_a-x_b$.} After this we are left with a single propagator connecting $x_k$ with an integration point $y_k$, which now can be treated with $\Box_k$.

It is easy now to figure out the general procedure. Assuming again that at the site $k-1$ there is no magnon we begin by acting with 
$\Box_{k-1}$, removing integration in $y_k$. Then, if at the site $k$ there is a magnon, we act with the combination $\Box_k x_{k,k-1}^2$, see figure \ref{magnon_move2}. Otherwise, on the sites with no magnon we act with the operator $x_{k,k-1}^2\Box_k$.
Acting in this way we can invert $\hat\cB$ completely.
For example for $J=3,M=1$ with $I_3=0$, $I_2=1$ and $I_1=0$ we find 
\beq\la{Binvex}
\hat\cB^{-1}=\frac{1}{(-4\xi^2)^3}x_{1,3}^2\,(x_{2,3}^2\Box_{3})\,(\Box_{2}x_{2,1}^2)\,\Box_{1}\;.
\eeq

We conclude that it is always possible to write $\hat\cB^{-1}$ as a differential operator. Below we rewrite $\hat\cB^{-1}$ in a simple and manifestly conformally invariant way, applicable for all cases. For that we will have to first change to embedding coordinates, introduced in section~\ref{secFC}.

\paragraph{Embedding coordinates.}
When manipulating conformal integrals such as the ones that are generated by the graph building operator, it is always advantageous to work in the embedding coordinates. 
In the case of the correlator (\ref{wf}), we have the $\chi_I$ operators in (\ref{chis}) at positions $X_{1},\dots,X_{J}$ and $\cO_\Delta$ at $X_0$. Hence, the corresponding embedding space wave function is defined as
\beq\la{embeddingwf}
\Phi^{\bf I}_\cO(X_1,\dots,X_J)\equiv{\varphi^{\bf I}_\cO(\vec X_1/X_1^+,\dots,\vec X_J/X_J^+)\over (X_1^+)^{1+I_1}\dots (X_J^+)^{1+I_J}\,(X_0^+)^{\Delta_\cO}}\;.
\eeq

The advantage of the covariant formalism is that the inverted $\hat\cB$ can be written in a very concise form
\beq\la{Binv1}
\hat\cB^{-1}=\tr\(\beta_{J,I_J}.\dots.\beta_{1,I_1}\)
\eeq
where $\beta_{k,I_k}$ is a local operator that only acts on the $X_k$'th coordinate. It is given by
\beq\la{Binv2}
\(\beta_{k,I_k}\)^{M\,N}=-{1\over2}\times\left\{\begin{split}X_k^MX_k^N\,K_k^2&\qquad I_k=0
\\
X_k^M\,K_k^2\,X_k^N&\qquad I_k=1
\end{split}\right.\ ,\qquad\text{where}\qquad K^M=-{i\over\xi}{\d\over\d X_M}\;.
\eeq
To understand the mechanism behind \eq{Binv1} we look at the same example as in \eq{Binvex}. There we took $I_3=0$, $I_2=1$ and $I_1=0$. For this case \eq{Binv1} gives
\beq
\cB^{-1}=\frac{1}{(-2)^3} (X_3.X_1) (X_3.X_2) K_3^2
K_2^2 (X_2.X_1) K_1^2\;.
\eeq
By noticing that $-\xi^2 K_i^2=\Box_i^{4D}-4\d_{X_i^+}\d_{X_i^-}$ and due to the fact that $\Phi^{\bf I}_{\cal O}$ from \eq{embeddingwf}
does not depend on $X_-$, we can simply replace $K_i^2$ by $-\frac{1}{\xi^2}\Box_i$ when acting on the CFT wave function. In addition 
we use $x_{ab}^2=\frac{-2X_a.X_b}{X^+_aX^+_b}$ to reproduce precisely \eq{Binvex}. It is equally easy to see that the equation \eq{Binv1} works in the general situation, provided that $J_2<J_1$.

Finally, let us rewrite the local site operators, $\beta_{k,I_k}$, in terms of the local generator of the conformal symmetry, namely in terms of ${\mathfrak q}_k$ defined as
\beq
\hat{\mathfrak q}_i^{M\,N}\equiv  X_i^N K_i^M-  X_i^M K_i^N\;.
\eeq
We can use that the CFT wave function
is a homogeneous function of $X_k$ with degree of homogeneity $-(1+I_k)$, as follows directly from its definition \eq{embeddingwf}.
This property allows us to make the replacements
$\hat X_k.\hat K_k\to {i\over\xi}(1+I_k)$.
Using that in analogy with \eq{qno}, we find
\beq\la{betatoq}
\beta_{j,I_j}^{MN}\simeq\(\hat {\mathfrak q}^2_k\)^{M\,N}-{i\over\xi}(2+I_k)\,{\mathfrak q}_k^{M\,N}-{1\over\xi^2}(1+I_k)\eta^{MN}\;.
\eeq
where the "$\simeq$" means equality when acting on the wave function (\ref{embeddingwf}).

\paragraph{The strong coupling classical limit.}
Before we proceed to the holographic interpretation of equation \eq{Binv1} at the quantum level,
we can already make a nontrivial conclusion about the strong coupling limit of the theory,
where the dual description becomes classical \cite{Gromov:2019aku}.

In order to understand the classical limit we recall that the operator $({\cal B}^{-1}-1)$
can be interpreted as a worldline Hamiltonian. The classical limit is obtained by simply replacing $-\frac{i}{\xi}\d_{x_i}$
by the classical momentum $p_i$~\cite{Gromov:2019aku}. As one can see from e.g. \eq{Binv2}
the only difference between the situation with magnon and without magnon is the ordering of the operators $X_i$
and $K_i$, which becomes irrelevant in the classical limit.
From this simple observation we conclude that
the classical theory is not affected by the presence of magnons.
But there will be a difference visible already at the first quasi-classical correction.

One way to think about this is that in the strong coupling limit the difference between a dimension one scalar and a dimension two composite operator made of two scalars (which is the magnon), becomes negligible. Namely, in units of $1/\xi$, both of them are indistinguishable from zero.

\subsection{Quantum fishchain with magnons}
In the previous section we understood that the classical limit of the theory with or without magnons should be the same.
This means that essentially all the quantization procedure developed in \cite{Gromov:2019bsj}
and described shortly in section~\ref{secFC}
still applies here.
There are only two steps in our quantization procedure which allow for a modification.
First, in \eq{Hqq} we assumed that the local Casimir operator $\tr \hat q^2_i$, which is a c-number, should be
associated with the dimension of the scalar, or more precisely its $AdS_5$ mass $m^2=-3$.
This assumption does not look natural anymore as the magnons are the bound-states of two scalars $\phi_1^\dagger$ and $\phi_2^\dagger$ of the total dimension $\Delta_i=2$, meaning that the $AdS_5$ mass becomes $-4$ or equivalently $\tr \hat q^2_i = \frac{8}{\xi^2}$.

Another subtle point in the quantization procedure, which needs to be revisited, is the ordering ambiguity in the quantum Hamiltonian $\hat H_q$ (see \eq{Handnq} and \eq{qno}). 
We have a freedom to add to the naive term $(\hat q_i^2)^{NM}$ the extra terms $(\hat q_i)^{NM}$ and $\eta^{NM}\tr q_i^2\sim \eta^{NM}/\xi^2$ like in \eq{qno}.
The general ansatz is
\beq\la{ambiguity}
(q_k^2)^{NM}\to  
:(q_k^2)^{NM}:_{I_k}=
(\hat q^2_k)^{NM}-i\frac{c_{I_k}}{\xi}\hat q_k^{NM}-\frac{d_{I_k}}{\xi^2}\eta^{NM}\;,
\eeq
where $c_{{I_k}}$ and $d_{I_k}$ are some numbers, which for the case with no magnons 
were set to $2$ and $1$ correspondingly i.e. $c_0=2,\;d_0=1$.
In this section we fix this quantization ambiguity by requiring that this quantum fishchain model is exactly dual to the Fishnet model for the states with magnons.

\paragraph{Map between the wave functions.}
In \cite{Gromov:2019bsj} we have built an explicit map between the CFT wave function in the ${\mathfrak u}(1)$-sector and the wave function of the quantum fishchain in $AdS_5$. This construction was reviewed in section~\ref{secFC}, see (\ref{bwf}). We now extend this map to the case with magnons.

The map \eq{bwf} takes the 4D flat space CFT wave function and returns a function of $J$ points $Z_1,\dots,Z_J$, parametrizing a unit radius $AdS_5$ space, $Z_i^2=-1$. These are the coordinates of the particles constituting the fishchain.

The factors $1/(Z_i.X_i)^3$ in (\ref{bwf}) are bulk to boundary propagators for $J$ scalars in $AdS_5$ of mass $m^2=-3$.
It was designed to solve the constraint 
$\tr \hat q^2_i= \frac{6}{\xi^2}$. As for the sites with magnons
we should modify this constraint to $\tr \hat q^2_i= \frac{8}{\xi^2}$, so we can take $1/(Z_i.X_i)^2$.
Hence, the generalization of the map (\ref{bwf}) to wavefunctions with both $\chi_0$ and $\chi_1$ operators is 
\beq\la{bwf2}
\psi_{\cal O}^{\bf I}(Z_1,\dots,Z_J)=
\int \prod_{i=1}^J \frac{D^4 X_i}{-4\pi^2(Z_i.X_i)^{3-I_i}}
\Phi_\cO^{\bf I}(X_1,\dots,X_J)\;.
\eeq
As before, $1/(Z_i.X_i)^2$ is the bulk to boundary propagator for an $AdS_5$ scalar of $m^2=-4$. It is the fastest decaying solution to the bulk equation of motion with a source at the boundary point $X_i$.\footnote{The other solution is  $\log(Z_i.X_i)/(Z_i.X_i)^2$.} 

It is not too hard to invert the map (\ref{bwf2}). One way is to go to the boundary of $AdS_5$. While with no magnon excitations the bulk wave function exactly reduces to the CFT wave function~\cite{Gromov:2019bsj}, in the case with magnons there is an addition logarithmic factor
\beq
\lim_{Z_i^+\to\infty}
\frac{
\psi_\cO^{\bf I}(Z_1,\dots,Z_J)}
{\prod_i(2\log Z_i^+)^{I_i}}
=\Phi_\cO^{\bf I}(Z_1,\dots,Z_J)\ .
\eeq

\paragraph{Exact duality.} Having the map
between the wave functions
extended to the general case
\eq{bwf2},
we can prove the duality and also fix the remaining 
quantization ambiguity manifested by the existence of two 
arbitrary constants in the QFC Hamiltonian \eq{ambiguity}.
Following~\cite{Gromov:2019bsj} we can establish the map between
the QFC Hamiltonian $\hat H_q$ and the CFT Hamiltonian $H$ 
by requiring 
\beq
\hat H_q\circ \psi_{\cal O}^{\bf I}(Z_1,\dots,Z_J)=
\int \prod_{i=1}^J \frac{D^4 X_i}{-4\pi^2(Z_i.X_i)^{3-I_i}}
H\circ\Phi_\cO^{\bf I}(X_1,\dots,X_J)=0\;,
\eeq
where $H=(\hat\cB^{-1}-1)$ as defined in (\ref{phys}). 
Exactly like in~\cite{Gromov:2019bsj} we simply use the identity
\beq
\int \prod_{i=1}^J \frac{D^4 X_i}{-4\pi^2(Z_i.X_i)^{3-I_i}}
\, \hat{\mathfrak q}_i\, F(X_1,\dots,X_J)=
\hat q_i\,\int \prod_{i=1}^J \frac{D^4 X_i}{-4\pi^2(Z_i.X_i)^{3-I_i}}
\,  F(X_1,\dots,X_J)\;,
\eeq
where the $q$'s are the $SO(1,5)$ isometry generators of $AdS_5$, (\ref{qZ}). 
From this 
we immediately establish the values of the missing constants in
\eq{ambiguity}
by reading them off \eq{betatoq}: $c_{I_k}=2+I_k$
and $d_{I_k}=1+I_k$.
Thus we arrive at the fishchain quantum Hamiltonian constraint
\beq\la{lift}
{\hat H}_q=\tr\[{:\hat{ q}_J:_{I_J}\over2}\dots{:\hat{ q}_1:_{I_1}\over2}\]-1\ .
\eeq
Here, $:{ q}_k:_{I_k}$ is given by 
\beq
\ \(:q^2_k\!:\!_{I_k}\)^{M\,N}\equiv
\( q^2_k\)^{M\,N}-{i\over\xi}(2+I_k)\,q_k^{M\,N}-{1\over\xi^2}(1+I_k)\eta^{MN}\;.
\eeq
This derivation extends the proof~\cite{Gromov:2019bsj} of the exact duality between the QFC and the Fishnet CFT models beyond the ${\mathfrak u}(1)$ sector.

Note that so far we only considered the CFT wave functions,
which contained $\phi_1^\dagger$ and $\phi_1^\dagger\phi_2^\dagger$ under the correlator with a local operator ${\cal O}$. Whereas this does indeed go beyond
the
${\mathfrak u}(1)$ sector
one could still question if that is the most general case. In particular one can also introduce ``anti-magnons" i.e. $\phi_2\phi_1^\dagger$, as we do in the next subsection. Later, however, we show that the CFT wave functions with the same magnon number $J_2$ (defined as a difference of the number of magnons and anti-magnons) are all equivalent to each other. It is yet very fruitful to have both magnons and anti-magnons at the same time as we will see in the next section \ref{sec:rep}.

\subsection{Adding anti-magnons to the QFC}\la{sec:rep}

We consider the most general situation, 
where at a site of the chain we can have either no magnon, one magnon, one anti-magnon or a magnon--anti-magnon pair. The CFT wave function
definition (\ref{wf}) should be extended by allowing for $\chi_j$ to take two additional values
\beq
{\includegraphics[scale=0.6]{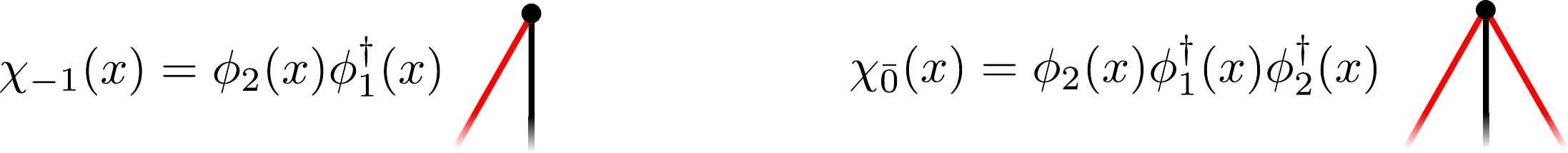}}\ .
\eeq
for anti-magnon and magnon--anti-magnon pair correspondingly. All the steps from the previous parts of this section are carried out in a similar way. We summarize in table~\ref{tab:summary} the key 
elements.
\begin{table}[h]
    \centering
    \begin{tabular}{|l|c|c|l|c|c|c|c|c|}\hline
    {type}& $I_k$ &  $\chi_{I_k}$  &$-2\beta_{I_k}^{MN}$ & $c_{I_k}$& $d_{I_k}$ &  $\Delta_k$ & $\tr q_k^2$ \\ \hline
    {no magnon}&  $0$ & $\phi_1^\dagger$ & $X^M X^N K^2$ & 2 & 1 & 1 & $6/\xi^2$\\
    {magnon}&  $+1$ & $\phi^\dagger_1{\color{red}\phi^\dagger_2}$& $X^M K^2 X^N$ & 3 & 2 & 2 & $8/\xi^2$\\
    {anti-magnon}&  $-1$&$
    {\color{red}\phi_2}\phi^\dagger_1$
    & $X^N K^2 X^M$ & 1 & 0 & 2 & $8/\xi^2$\\
    {magnon + anti-magnon}&  $\bar 0$ & 
    ${\color{red}\phi_2}\phi^\dagger_1{\color{red}\phi^\dagger_2}$
    & $K^2 X^M X^N$ & 2 & 1 & 3 & $6/\xi^2$\\ \hline
    \end{tabular}
    \caption{Four types of local composite operators entering into the definition of the CFT wave function.}
    \label{tab:summary}
\end{table}

As before we assume that $|M-\bar M|<J$ where $M$ is the total number of magnons and $\bar M$
is the total number of anti-magnons. In this case we again have  $\hat{\cB}^{-1}=\tr(\beta_{J,I_J}.\dots.\beta_{1,I_1})$ where the $\beta$'s are summarized in table~\ref{tab:summary}.

Next, one can also map the CFT wave function with the QFC wave function in $AdS_5$ using the generalization of
\eq{bwf2}
\beq\la{bulkwf}
\Psi^{\bf I}(Z_1,\dots,Z_J)
= \int\prod {D^4X_i\over-4\pi^2}{\Phi^{\bf I}(X_1,\dots,X_J)\over(X_1.Z_1)^{4-\Delta_1}\dots(X_J.Z_J)^{4-\Delta_J}}
\eeq
where again $\Delta_k$ is the conformal dimensions of $\chi_{I_k}$, given in the table~\ref{tab:summary}. Finally, the QFC Hamiltonian $\hat H_q$ is obtained by uplifting the
CFT Hamiltonian $\hat H$, which results in fixing the quantization ambiguity parameters $c_I$ and $d_I$
as shown in the table~\ref{tab:summary}.

To summarize, we see that the quantum fishchain naturally generalizes to incorporate magnons. They are in one to one correspondence with all possible orderings of $K^2_k$ and the $X_k$'s in the site operator $\beta_k$. 

\section{Integrability of the fishchain with magnons}\la{sec4} 
In this section we extend the proof of the quantum integrability of the QFC~\cite{Gromov:2019bsj} to the general case with magnons. In order to demonstrate the integrability of the QFC model we will follow the same steps as in \cite{Gromov:2019bsj}. 
The strategy is to identify the Hamiltonian as a part of a large family of mutually commuting operators, which can be obtained as an expansion in powers of $u$ of the T-operators $\hat {\mathbb T}(u)$.

There are $3$ non-trivial T-operators,
which should contain a complete set of integrals of motion.
The building blocks for these operators are the Lax matrices in antisymmetric finite-dimensional representations of $\mathfrak{sl}(4)$. They were computed in \cite{Gromov:2019bsj} in terms of $\hat q_k$. We will see that this derivation is general enough
to accommodate the current case as well.
The main building blocks for the operators $\hat{\mathbb T}(u)$ are the L-matrices
\beqa\la{generalL}
\hat {\mathbb L}_k^{{\bf 1}}(u)&=&1\\ \nn
\hat {\mathbb L}_k^{\bf 4}(u)&=&u
-\tfrac{i}{2}\hat q_k^{MN}\Sigma_{MN}\\ \nn
\hat {\mathbb L}_k^{\bf 6}(u)&=&u^2+u \hat q_k+\frac{\hat q_k^2}{2}
-\frac{i\hat q_k}{\xi}
-\frac{\tr \hat q_k^2}{8}+\frac{1}{4\xi^2}
\\ \nn
\hat {\mathbb L}_k^{\bar{\bf 4}}(u)&=&
\(u^2-\frac{1}{8}\tr\hat q_k^2+\frac{1}{\xi^2}\)\[-\hat {\mathbb L}_k^{\bf 4}(-u)\]^T\\ \nn
\hat {\mathbb L}_k^{\bar{\bf 1}}(u)&=&\(u^2-\frac 18 \tr\hat q_k^2+\frac{5}{4\xi^2}\)^2+\frac 1{8\xi^2} \tr\hat q_k^2-\frac{1}{\xi^4}\;.
\eeqa
Where $\Sigma_{MN}$ are $6$-dimensional $\sigma$-matrices.
Their explicit representation can be found in~\cite{Gromov:2019bsj}.
The mutually commuting families of operators are constructed from \eq{generalL} via
\beqa\la{LtoT}
\hat {\mathbb T}(u)=\tr\!\! \[
\hat {\mathbb L}_J(u-\theta_J)
\dots \hat {\mathbb L}_{2}(u-\theta_2)
\hat {\mathbb L}_{1}(u-\theta_1)G
\]\;.
\eeqa
Where $G$ is a constant {\it twist} matrix and $\{\theta_j\}$ are the so-called inhomogeneities which can be taken as arbitrary.
For the original fishnet model the twist matrix is trivial, $G=1$. However,
here we consider the most general deformed case, dual to the twisted fishnet of~\cite{twistingpaper},
with $G={\rm diag}(\lambda_1,\lambda_2,\lambda_3,\lambda_4)$ and we assume $\lambda_1\lambda_2\lambda_3\lambda_4=1$ for unimodularity of $G$. Next, we have to choose the values of $\{\theta_j\}$ so that the QFC Hamiltonian 
appears in \eq{LtoT}  for a special value of $u$. More precisely we
require that in the same manner as in the case without magnons
we have 
$\hat {\mathbb T}^{\bf 6}(0)=\hat H_q+1$. 

We now consider all $4$ cases of insertions one by one -- the site without magnon, with magnon, with anti-magnon, and with magnon anti-magnon pair using the data from table~\ref{tab:summary}.

\paragraph{No-magnon case.} The case without magnon corresponds to a single scalar $\phi_1$, which allows for the value of the quandatic Casimir to be $-3$, implying $\tr \hat q_k^2=\frac{6}{\xi^2}$. As a result \eq{generalL} reduces to
\beqa\la{generalL0}
\hat {\mathbb L}_k^{\bf 6}(u)&=&u^2+u \hat q_k+\frac{\hat q_k^2}{2}
-\frac{i\hat q_k}{\xi}
-\frac{1}{2\xi^2}
\\ \nn
\hat {\mathbb L}_k^{\bar{\bf 4}}(u)&=&
\(u^2+\frac{1}{4\xi^2}\)\[-\hat {\mathbb L}_k^{\bf 4}(-u)\]^T\\ \nn
\hat {\mathbb L}_k^{\bar{\bf 1}}(u)&=&u^2\(u^2+\frac{1}{\xi^2}\) \;,
\eeqa
we see that $\hat {\mathbb L}_k^{\bf 6}(0)=\frac{\hat q_k^2}{2}
-\frac{i\hat q_k}{\xi}
-\frac{1}{2\xi^2}=\frac{:\hat q_k^2:_0}{2}$, which is the correct factor in $\hat H_q$ for sites without a magnon, meaning that for these sites we have to set $\theta_k=0$.

\paragraph{Magnon case.}
For the site with magnon the dimension $\Delta_k=2$ corresponds 
to the quadratic Casimir $-4$, implying that $\tr \hat q_k^2=\frac{8}{\xi^2}$. Consequently from \eq{generalL} we get
\beqa\la{generalL1}
\hat {\mathbb L}_k^{\bf 6}(u)&=&u^2+u \hat q_k+\frac{\hat q_k^2}{2}
-\frac{i\hat q_k}{\xi}
-\frac{3}{4\xi^2}
\\ \nn
\hat {\mathbb L}_k^{\bar{\bf 4}}(u)&=&
u^2\[-\hat {\mathbb L}_k^{\bf 4}(-u)\]^T\\ \nn
\hat {\mathbb L}_k^{\bar{\bf 1}}(u)&=&\(u+\frac{i}{2\xi}\)^2\(u-\frac{i}{2\xi}\)^2\;.
\eeqa
This time the ambiguity constants are $c_I=3$ and $d_I=2$ i.e. we need to tune $\theta_k$ to obtain $\hat {\mathbb L}_k^{\bf 6}(-\theta_k)=\frac{\hat q_k^2}{2}
-\frac{3i\hat q_k}{2\xi}
-\frac{1}{\xi^2}$. Note that with only one parameter $\theta_k$ it is not guaranteed that we manage to fit the two constants correctly. Fortunately, by setting $\theta_k=\frac{i}{2\xi}$\footnote{N.G. is grateful to V.Kazakov and Z.Tsuboi for discussing the possibility of inclusion of inhomogeneities in relation to the magnons.}
we indeed obtain 
$\hat {\mathbb L}_k^{\bf 6}(0)=\frac{1}{2}:\hat q_k^2:_1$.
\paragraph{Anti-magnon case.}
The building block for the anti-magnon is
$\hat {\mathbb L}_k^{\bf 6}(-\theta_k)=\frac{\hat q_k^2}{2}
-\frac{i\hat q_k}{2\xi}
-\frac{1}{\xi^2}$, which corresponds to $\theta_k=-\frac{i}{2\xi}$.

\paragraph{Magnon--anti-magnon case.} The Lax matrix, in this case, is identical to the no-magnon case considered above. 
The reason is that $\Delta(\chi_{\bar 0})=4-\Delta(\chi_0)=3$, so the quadratic Casimir, given by $c=\Delta(\Delta-4)=-3$ is exactly the same as that for $\Delta=1$. This implies that in both cases we have $\tr q_k^2=\frac{6}{\xi^2}$. Moreover, the quantization ambiguity constants $c_I$ and $d_I$ are also exactly the same  (see table~\ref{tab:summary}) and therefore $\theta_k=0$ in this case as well. As is explained in the next section, at the fishnet side both the magnon--anti-magnon case and no-magnon case describe different ways of cutting the same propagation line of $\phi_1$ along the fishnet diagrams. Hence, the corresponding transfer matrix that encodes all physical charges should be the same. 

To summarize, we see that after fixing the impurities to $\theta_k=\frac{iI_k}{2\xi}$ we get indeed 
$\hat {\mathbb T}^{\bf 6}(0)=\hat H_q+1$.
This property implies the quantum integrability of our model, as we identified the Hamiltonian of the system within a large family of
mutually commuting operators.\footnote{Of course one should verify that there are sufficiently many independent integrals of motion. This is usually rather hard to verify rigorously, but we strongly believe this is the case.
The rough counting of the integrals of motions indeed gives $\sim 4J$, which coincides with the number of degrees of freedom of the theory. }

In the current formulation the ordering of the magnons is quite important.
Different distributions of magnons within the chain will produce different
Hamiltonian and T-operators. 
We will see in section~\ref{sec:integ2} that the eigenvalues of all T-operators are in fact independent of the order of magnons. This invariance has a very nice physical interpretation -- it can be associated with the discrete reparametrization symmetry of the fishchain as we discuss in the next section~\ref{sec5}.
Based on the results of this section, in section~\ref{sec:integ2}
we present the general integrability formalism for the spectrum of the theory. 

\section{Discrete reparametrization symmetry}\la{sec5}
When introducing the magnons we have to face the following paradox. On the CFT side, the primary operators with several magnons will be given by a complicated linear combination of all possible magnon permutations. At the same time on the fishnet side, it looks like we have the freedom to choose any ordering for the magnons. This choice is made explicitly in our definition of the CFT wave function by the correlator (\ref{U2}). Moreover, the fishchain Hamiltonian $\hat H_q$ itself depends on a particular fixed order. The danger is that the spectrum of the Hamiltonian is also order dependent, which would result in an inconsistency of our approach. The physical quantities such as dimensions of local operators should not depend on an arbitrary choice of order of insertions of $\phi_2$ fields in the CFT wave function, which probes this local operator.
\begin{figure}[t]
\centering
{\includegraphics[scale=0.7]{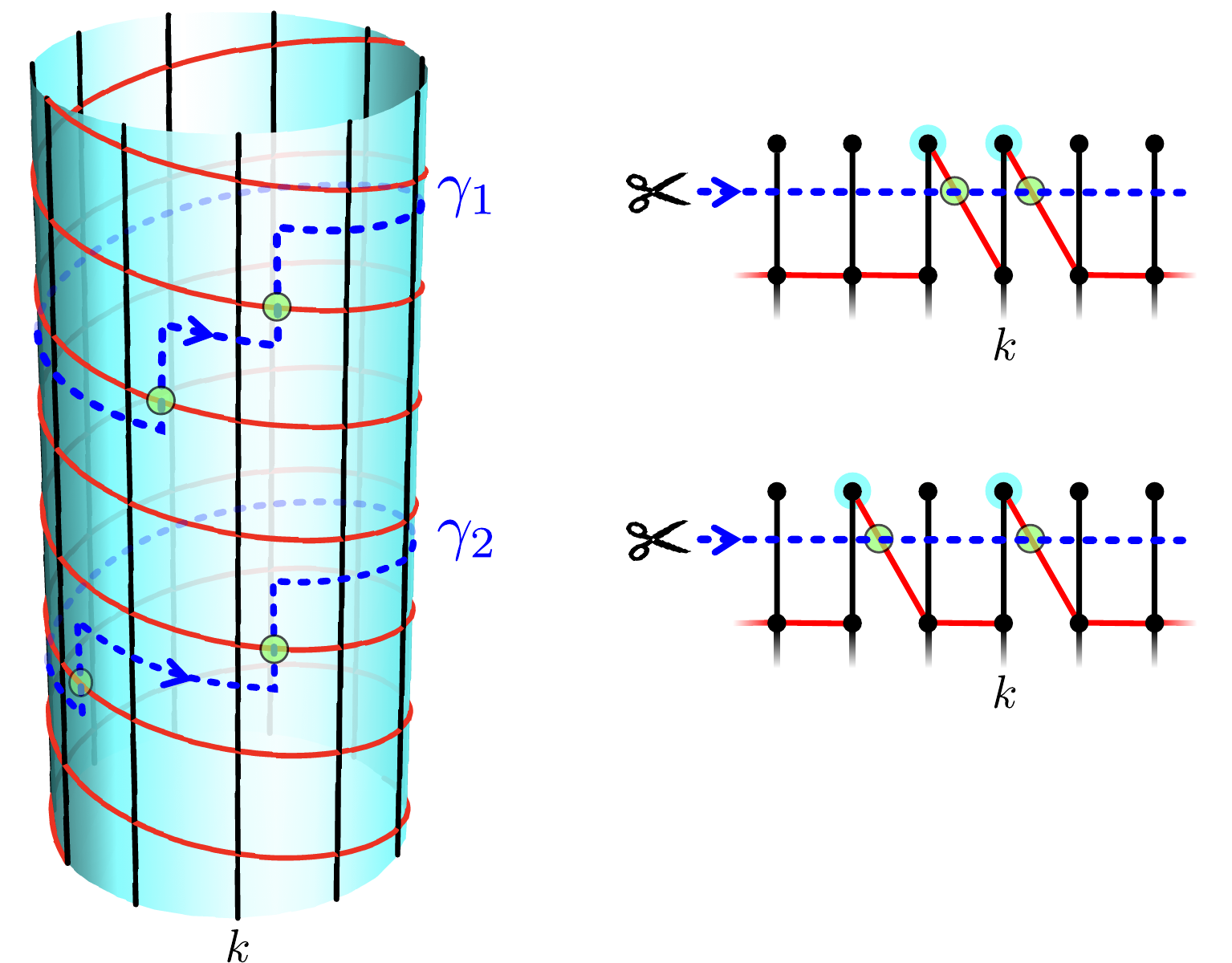}}
\caption{Two possible cuttings of the two point function of two local operators ${\cal O}$. Each cutting represents the same two point function as a contraction of two CFT wave functions. The two cuts produce different ordering of the magnons inside the chain. The cut $\gamma_1$ leads to the wave function (\ref{wf}) with $\{I_{k+3},I_{k+2},I_{k+1},I_k,I_{k-1},I_{k-2}\}=\{0,0,1,1,0,0\}$, while the cut $\gamma_2$ leads to the equivalent wave function with $\{I_{k+3},I_{k+2},I_{k+1},I_k,I_{k-1},I_{k-2}\}=\{0,1,0,1,0,0\}$.}\label{cuts}
\end{figure}

In this section we demonstrate that, luckily,
the spectrum does not depend on the ordering of the magnons. This is due to a novel symmetry of the fishchain which we interpret as the {\it discrete reparametrization} symmetry.
We will show that redistribution of the magnons can be understood as a map between Hilbert spaces of the two chains, which differ by the magnon positions. We will then prove that this map preserves the eigenvalues of $\hat H_q$ and of all other
conserved charges encoded into $\hat{\mathbb T}(u)$, introduced in the previous section.

\subsection{Cutting and leveling}
Consider a correlation function between two local operators of the type (\ref{U2}) that consists of $J_1=J$ fields $\phi_1$ and $J_2=M$ insertions of $\phi_2$'s fields in the trace (as before we assume $M<J$). The planar Feynman diagrams that contribute to this correlator have cylindrical topology and are of an iterative spider-net-like structure as discussed in the introduction, see figure \ref{cuts}. 
The fishchain state, as given by the correlator in (\ref{wf}), corresponds to a specific choice of cut around this cylinder. For example the blue cut $\color{blue}\gamma_1$ in figure \ref{cuts} corresponds to the correlator (\ref{wf}) with $M=2$, where one magnon ($\chi_1$) is at the 
$k$'th site of the chain and the second magnon is at the neighboring $k+1$ site. On the other hand, when the same diagram is cut as is indicated by the $\color{blue}\gamma_2$ cut in figure \ref{cuts}, the second magnon is at the $k+2$ site. 

In general, a cut is a choice of a decomposition of the fishnet diagrams. It is specified by a line ${\color{blue}\gamma}$ that goes around the cylinder and crosses a set of propagators on the way. Here, we choose to restrict our consideration to {\it space-like cuts}. These are cuts that cross every $\phi_1$ line and every $\phi_2$ line exactly once. Moreover, every $\phi_2$ cut must be followed by a $\phi_1$ cut. These are the cuts that look space-like on the fishnet lattice. They are the ones that allow us to obtain all the higher loop fishnet diagrams by iterative action of the corresponding graph building operator. Such cuts are analogues to Cauchy surfaces on the string worldsheet. The graph building operator generates the ``worldsheet'' time propagation on such surfaces.

We notice that a choice of space-like cut corresponds to a choice of a basis of CFT wave functions of the type (\ref{U2}) as follows. A cut of a $\phi_1$-propagator is mapped to a $\phi_1^\dagger$ field in the corresponding correlator (\ref{U2}) and a cut of a $\phi_2$-propagator is mapped to a $\phi_2^\dagger$ field (see figure~\ref{cuts}). We then merge $\phi_1^\dagger$ and $\phi_2^\dagger$
fields to be at the same point as a $\chi_1$ or $\chi_{-1}$ in accordance with the vertex above the cut.
We will give more explicit examples below.

\begin{figure}[h]
\centering
\def\svgwidth{13.5cm}

\ifpdf
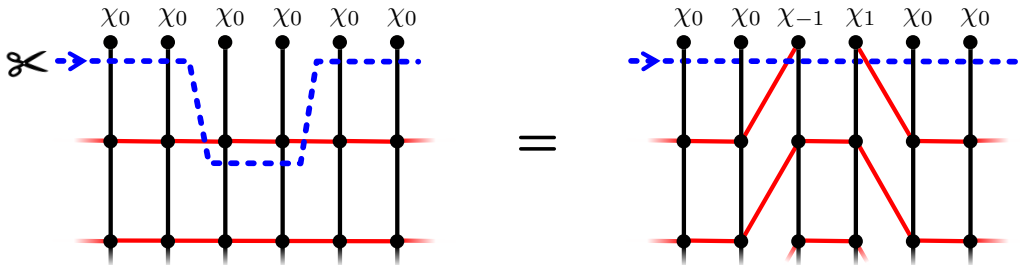
\else
\com{PDF-picture replacement}
\fi

\caption{ Another example of a cut which, instead of moving magnons around, creates a magnon--anti-magnon pair.}\label{cut_deformation}
\end{figure}

Let us point out that the situation in which two different cuts result in
a different order of magnons in the chain like in
 figure~\ref{cuts} is not the most general.
In fact one can also generate an arbitrary number of
magnon--anti-magnon pairs. For example, we can start with a chain with no magnons at all and deform the cut. After doing so, we end up with a magnon and an anti-magnon, see figure \ref{cut_deformation}. 

One way to think about the difference between different cuts is as a 
map between Hilbert spaces. We will now construct explicit operators that implement this map between the Hilbert spaces of two different deformations of the cut.

\subsection{The cut deformation operators}

Any deformation of the cut, such as the one in figure \ref{magnon_move2}, can be realized as the action of a {\it cut deformation operator} on the corresponding CFT wave function. The aim of this section is to construct the CFT cut deformation operators and then lift them into
an operator acting on fishchain wavefunctions, i.e. acting on the $AdS_5$ variables $Z_i$. 
We will use these operators in section \ref{sec:integ2} and will prove that they generate a discrete symmetry of the quantum fishchain, leaving the spectrum invariant.

\begin{figure}[h]
\centering
\def\svgwidth{14cm}

\ifpdf
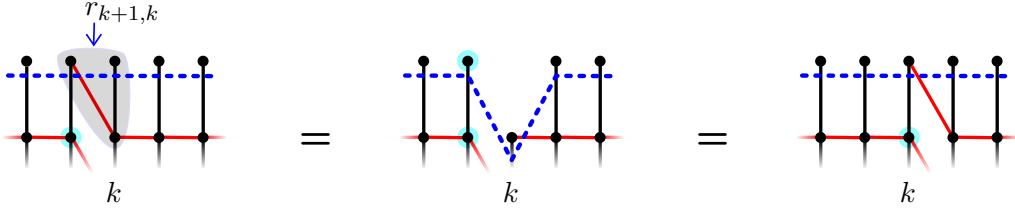
\else
\com{PDF-picture replacement}
\fi

\caption{\small The operator $r_{k+1,k}$ moves a magnon at the $(k+1)$ site to the empty site on its right, ($k$). Similarly, the reflection of this figure describes 
how the operator $\bar r_{k+1,k}$ moves an anti-magnon at the $k$'th site to the empty site on its left ($k+1$).}\label{basic_moves6}
\end{figure}

Any cut deformation of the CFT wave function can be decomposed into a combination of two elementary operations
\begin{figure}[h]
\centering
\def\svgwidth{12cm}

\ifpdf
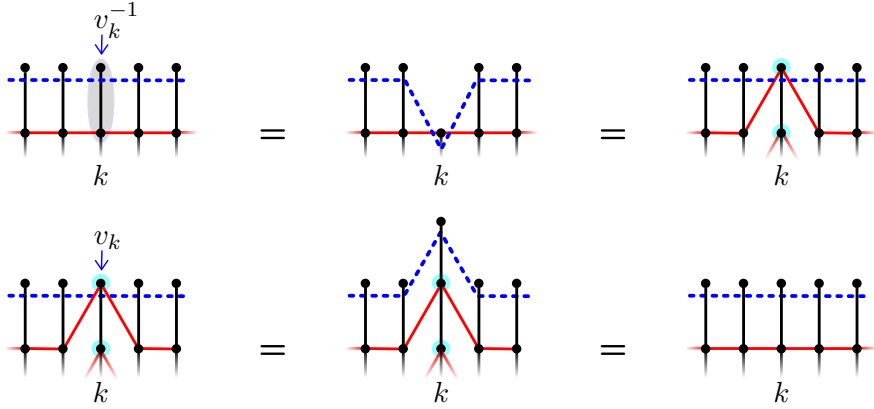
\else
\com{PDF-picture replacement}
\fi

\caption{Graphical representation of the action of the operators $v_k^{-1}$ and $v_k$ on a CFT wave function. The operator $v_k^{-1}$ removes a vertical $\phi_1$ propagator, which is equivalent to the creation of a magnon--anti-magnon pair on one site. Correspondingly, $v_k$ annihilates the magnon--anti-magnon pair.}\label{vact}
\end{figure}
\beqa\la{elementarymoves}
&v_k\circ\Psi(\dots,X_k,\dots)={\xi^2\over\pi^2}\int D^4Y_k\,\frac{\Psi(\dots,Y_k,\dots)}{-2X_k.Y_k}\;\;\;\;&\text{(add vertical propagator)}\\
&h_{k+1,k}\circ\Psi(\dots,X_{k+1},X_k,\dots)=
\frac{\Psi(\dots,X_{k+1},X_k,\dots)}{-2X_{k+1}.X_j}
\;\;\;\;&\text{(add horizontal propagator)}
\ .
\eeqa
As one can see from the equation the
$v_k$ operator adds a vertical $\phi_1$ propagator, and $h_k$ adds a horizontal $\phi_2$ propagator. The corresponding inverse of these elementary operations are
\begin{figure}[b]
\centering
\def\svgwidth{8cm}

\ifpdf
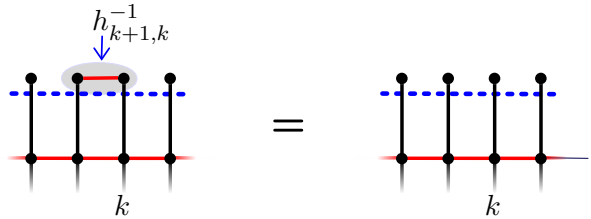
\else
\com{PDF-picture replacement}
\fi

\caption{\small Graphical representation of the action of the operator $h_{k+1,k}^{-1}$, removing the $\phi_2$ propagator stretched horizontally between the $k+1$ and $k$ sites. It can be interpreted as an operator annihilating a magnon and anti-magnon situated at two neighbouring sites.}\label{hact}
\end{figure}
\beqa\la{invem}
&v_k^{-1}=-{1\over4\xi^2}\,\Box_k\;\;\;\;&\text{(remove vertical propagator)}\\ &h_{k+1,k}^{-1}=-2X_{k+1}.X_k\;\;\;\;&\text{(remove horizontal propagator)}\ .
\eeqa
We can build out of $v$ and $h$
the operators which move a magnon/anti-magnon to the right/left
\beqa\la{Vinv}
&r_{k+1,k}\equiv h_{k+1,k}^{-1}\circ v_k^{-1}
\;\;\;\;&\text{(moves magnon right)}\\
&\bar r_{k+1,k}\equiv h_{k+1,k}^{-1}\circ v_{k+1}^{-1}
\;\;\;\;&\text{(moves anti-magnon left)}
\;.
\eeqa
Let us show that the operators $v_k,\;r_{k+1,k},\;\bar r_{k+1,k}$ and $h^{-1}_{k+1,k}$ constitute the $3$ elementary moves which allow us to map any CFT wave function to a standard form.
When acting on a state in a fishchain Hilbert space, characterized by the indexes ${\bf I}$, they transform
it to a state in a different Hilbert space, characterized by a different set of indexes $\tilde{\bf I}$ as follows
\beqa
v_k\;\;:\;\;{\bf I}=\{\dots I_{k+1},\bar 0,I_{k-1} \dots\}\;\;&\to&\;\;\tilde{\bf I}=\{\dots I_{k+1}, 0,I_{k-1} \dots\}\\
r_{k+1,k}\;\;:\;\;{\bf I}=\{\dots I_{k+2},1,0,I_{k-1} \dots\}\;\;&\to&\;\;\tilde{\bf I}=\{\dots I_{k+2},0,1,I_{k-1} \dots\}\\
\bar r_{k+1,k}\;\;:\;\;{\bf I}=\{\dots I_{k+2},0,-1,I_{k-1} \dots\}\;\;&\to&\;\;\tilde{\bf I}=\{\dots I_{k+2},-1,0,I_{k-1} \dots\}\\
h^{-1}_{k+1,k}\;\;:\;\;{\bf I}=\{\dots I_{k+2},1,-1,I_{k-1} \dots\}\;\;&\to&\;\;\tilde{\bf I}=\{\dots I_{k+2},0,0,I_{k-1} \dots\}\;,
\eeqa
see figures~\ref{hact} and \ref{basic_moves6}.
We are going to show that a state with ${\mathfrak u}(1)$ charges $|J_2|<J_1$ can be mapped to the state 
in the Hilbert space, characterized by ${\bf I}$ in the standard configuration
${\bf I}=\{1,1,\dots,1,0,0,\dots,0\}$, for $J_2\leq 0$,
and ${\bf I}=\{-1,-1,\dots,-1,0,0,\dots,0\}$ otherwise.
Indeed, we can use $v_k$ to replace all $\bar 0$ by $0$'s. Next we can move magnons and anti-magnons towards each other using $r_{k+1,k}$ and $\bar r_{k+1,k}$ and then annihilate them in pairs using $h^{-1}_{k+1,k}$.
After that, depending on the sign of $J_2$ there will be only magnons or anti-magnons left, which we can group together with $r_{k+1,k}$ or $\bar r_{k+1,k}$.

In the next section we derive what these operators correspond to in the dual description.

\subsection{The fishchain reparametrization operators}

Using the map between the CFT and the fishchain wavefunctions (\ref{bulkwf}) we now map $v_k^{-1}$ and $h_k^{-1}$ into fishchain operators. Consider first the operator $v_k^{-1}$. This operator acts on a site with no magnon and maps it to a site with a magnon anti-magnon pair. Plugging it into the map (\ref{bulkwf}) and taking into account that the conformal dimension at that site changes from $\Delta_k=1$ to $\Delta_k=3$, we get the equation for the corresponding operator ${\mathbb V}^{-1}$ in $AdS_5$
\beq\la{vlift}
v^{-1}(x)\circ\frac{1}{(Z.{\cal X})^1}={\mathbb V}^{-1}(Z)\circ
\frac{1}{(Z.{\cal X})^3}\;.
\eeq
Using that $v^{-1}(x)=
-{1\over4\xi^2}\Box_{\vec x}
$ we conclude that ${\mathbb V}^{-1}(Z)={1\over2\xi^2}$, i.e. 
up to a multiplication by a constant this operator acts trivially in the dual picture. This is because on the fishchain side there is no real distinction between no-magnon and
a magnon--anti-magnon pair at the same site, as one can see from the table~\ref{tab:summary}.

Similarly, for $h_{k+1,k}^{-1}$ (which decreases $\Delta_i$ by one on two consecutive sites) we find its $AdS_5$ counterpart ${\mathbb H}$ from the following identity
\beq
{\mathbb H}_{\Delta_1,\Delta_2}(Z_1,Z_2)\circ
\frac{1}{(Z_1.{\cal X}_1)^{4-\Delta_1}(Z_2.{\cal X}_2)^{4-\Delta_2}}
=
\frac{-2{\cal X}_1.{\cal X}_2}{(Z_1.{\cal X}_1)^{5-\Delta_1}(Z_2.{\cal X}_2)^{5-\Delta_2}}\;,
\eeq
where $\Delta_j$ can be either $2$ or $3$. 
We find that 
\beq\la{Hres}
{\mathbb H}_{\Delta_1,\Delta_2} (Z_1,Z_2)={-2\over\alpha_1\alpha_2}\[\nabla_1.\nabla_{2}+\alpha_1 Z_1. \nabla_{2}+\alpha_2Z_2.\nabla_{1}+\alpha_1\alpha_2Z_1.Z_2\]
\eeq
where $\alpha_k=4-\Delta_k$
and\footnote{Remember that the embedding coordinates should satisfy $Z^2=-1$, so the $\d_Z$ operation is in principle not very well defined, however the combination $\nabla$ commutes with $Z^2$, making it well defined.}
\beq\la{AdSderivative}
\nabla_{j,M}={\d\over\d Z^M_j}+Z_{j\,M}\(Z_j^N{\d\over\d{Z_j^N}}\)\;.
\eeq
As a consequence, the lift of the operator $r_{k+1,k}=h_k^{-1}\circ v_k^{-1}$, that moves a magnon to an empty site to the right reads, up to an irrelevant constant factor
\beq\la{Rop}
{\mathbb H}_{3,2}=
\nabla_{1}.\nabla_{2}+ 2Z_2.\nabla_{1}+Z_1.\nabla_{2}+2 Z_1.Z_2\;.
\eeq
In the next section we will show that the {\it magnon move} operator ${\mathbb H}_{3,2}$ and
{\it magnon annihilation} operator ${\mathbb H}_{2,2}$
generate a gauge symmetry of the fishchain that we interpret as discrete  reparametrization symmetry.

\section{Integrability, Exact Spectrum and Reparametrization Invariance}
\la{sec:integ2}

In section \ref{sec4} we constructed a set of mutually commuting operators which also include the Hamiltonian $\hat H_q$. Then, in section \ref{sec5}, we have constructed the fishnet cut deformation operators and lifted them to fishchain operators. In this section, we  use the set of commuting operators to solve for the spectrum of dimensions $\Delta$ of local operators. We  then show that the cut deformation operators generate similarity transformations of the transfer matrices that can be interpreted as gauge transformations of the corresponding Baxter equation.

This section generalizes the results of \cite{Gromov:2017cja}, where the non-perturbative spectrum was obtained for the operators $J=3,\;M=0$\footnote{We remind that $M$ denotes the number of magnons and $\bar M$ number of anti-magnons. The ${\mathfrak u}(1)$ charges in these notations are $J_1=J,\;J_2=M-\bar M$.}, and \cite{Gromov:2019cja}, where the general $J,\;M=0$ case was considered. An important result of \cite{Gromov:2017cja,Gromov:2019cja}, which we are going to use here is the {\it quantization condition}, which is needed in addition to the Baxter TQ-relations. We will propose the generalization of this quantization condition and also present the most general form of the TQ-relations valid for any values of the charges $J_1,J_2$ such that $J_1>|J_2|$.

Using these techniques we will be able to
solve numerically for several examples of states, $J=3,M=1$ and $J=3,M=2$. At finite $\xi$
we explore a very rich analytic structure of the function $\Delta(\xi)$. At weak coupling 
we are able to compare our numerical results
with the asymptotic Bethe ansatz, valid only up to
the wrapping order $\xi^{2J+2|M|-2}$.

Our Baxter TQ-relation, which we present in this section, together with the generalization of the quantization condition of~\cite{Gromov:2017cja,Gromov:2019cja} constitute the complete set of equations, analogous to the Quantum Spectral Curve equations for N=4 SYM~\cite{Gromov:2013pga} or ABJM~\cite{Cavaglia:2014exa} theory.

\subsection{T-functions and Baxter TQ relation}
We start by formulating the TQ-relations. In order to compare with the previous partial results of \cite{Gromov:2017cja,Gromov:2019cja}
it is convenient to rescale the spectral parameter by $\xi$. To avoid confusion we denote the rescaled spectral parameter by $v$, which is related to the initial one by $u=v/\xi$.\footnote{When writing ${\mathbb T}(v)$, strictly speaking, what we mean is ${\mathbb T}(v/\xi)$.} We also denote by ${\mathbb T}(u)$ the eigenvalue (or T-functions) of the simultaneously 
diagonalizable operators $\hat {\mathbb T}(u)$.

The T-functions corresponding to the trivial representations can be computed explicitly 
\beq
{\mathbb T}^{\bf 1}(v)=1\;\;{\rm and}
\;\; 
{\mathbb T}^{\bar {\bf 1}}(v)
=\frac{v^{2J}(v+i)^{J-M+\bar M}(v-i)^{J+M-\bar M}}{\xi^{4J}}\;.
\eeq
Also we notice from (\ref{LtoT})-\eq{generalL1} that ${\mathbb T}^{\bar{\bf 4}}$ and
${\mathbb T}^{{\bf 4}}$ are closely related
\beq\la{Tbar}
{\mathbb T}^{\bar{\bf 4}}(v)=
\frac{(-1)^J}{\xi^{2J}}(v+\tfrac i2)^{J-M+\bar M}
(v-\tfrac i2)^{J+M-\bar M}
{\mathbb T}^{{\bf 4}*}(-v)
\eeq
where $*$ indicates that one should inverse the order of the particles and interchange magnons and anti-magnons.
This transformation is an obvious symmetry of the system, but
of course not all states are invariant under this transformation.

\paragraph{Baxter TQ-relations.} One of the key quantities in 
quantum integrability are the Q-functions~\cite{Baxter:1972wg,Bazhanov:1996dr}. They can be defined as T-functions corresponding to a rather complicated auxiliary space representation. As a simple alternative to defining them as eigenvalues of these complicated Q-operators, one can find them from a finite difference TQ-relation, which in our case reads\footnote{We use the standard shorthand notations $f^{[+a]}=f(v+\tfrac{i a}{2})$.}
\beq\la{TQ}
{\mathbb T}^{{\bf 1}[+2]}\;Q^{[+4]}
-
{\mathbb T}^{{\bf 4}[+1]}\;Q^{[+2]}
+
{\mathbb T}^{{\bf 6}}\;Q
-
{\mathbb T}^{\bar {\bf 4}[-1]}\;Q^{[-2]}
+
{\mathbb T}^{\bar {\bf 1}[-2]}\;Q^{[-4]}=0\;.
\eeq
We notice that  the coefficients in this equation are polynomials of different degree.
However, as we can see from \eq{Tbar} 
even though ${\mathbb T}^{\bar{\bf 4}}(v)$
is a polynomial of degree $3J$, 
it has a trivial part of degree $2J$, which factorises. To remove the trivial factors we introduce the following {\it gauge} transformation
\beq
Q(v)=q(v)\; v^{\bar M}\[\xi^{i v }e^{\frac{\pi  v}2}\Gamma(-i v)\]^J\;.
\eeq
This transformation is designed so that the  equation \eq{TQ} becomes more symmetric
\beqa\la{BaxterGeneral0}
-\frac{P_{2J}^{\bf 6}(v)}{v^{J-M-\bar M}}\;q_a&=&v^M (v+i)^J (v+2i)^{\bar M}\; q_a^{[+4]}
-
v^M(v+i)^{\bar M}P_J^{\bf 4}(v+\tfrac i2)\;q_a^{[+2]}\\
\nn
&+&
v^{\bar M}(v-i)^{J}(v-2i)^M\;q_a^{[-4]}
-
v^{\bar M}(v-i)^M P_J^{\bar{\bf 4}}(v-\tfrac i2)\;q_a^{[-2]}
\;,
\eeqa
where we defined the degree-$L$ polynomials $P_L$ in the following way
\beq\la{PinT}
P_J^{\bf 4}(v)\equiv\xi^J {\mathbb T}^{\bf 4}(v)\;\;,\;\;
P_{2J}^{\bf 6}(v)\equiv\xi^{2J} {\mathbb T}^{\bf 6}(v)
\;\;,\;\;P_J^{\bar{\bf 4}}(v)\equiv(-\xi)^J {\mathbb T}^{\bf 4 *}(-v)\;.
\eeq
The index $a=1,\dots,4$ in $q_a$ labels $4$ 
independent solutions of \eq{BaxterGeneral0}.
We specify below exactly how these solutions
are defined by their large $v$ asymptotic.

One can see that in the form \eq{BaxterGeneral0} the Baxter TQ-equation generalizes the one derived in the CFT context in \cite{Gromov:2017cja,Gromov:2019cja}
for the case $M=0$. Furthermore, in
\cite{Gromov:2017cja} it is in addition assumed that the state is invariant under the inversion of the order of the sites, which we do not assume here.
Our equation is thus a generalization of the results of \cite{Gromov:2017cja,Gromov:2019cja} to the most general states
with an arbitrary number of magnons, as long as $|J_2|<|J_1|$.

\paragraph{Asymptotics of q.}
The global conformal group charges $\Delta,S_1$ and $S_2$,
enter into the TQ-relations through the
large $v$ asymptotic of T-functions and consequently Q-functions.
Indeed, at large $v$ the product \eq{LtoT}
becomes a sum of $\hat q_k$, which is the global $\mathfrak{so}(1,5)$ charge.
This observation is reflected in the following expressions for the highest powers of the polynomials $P_L$:
\beqa\nn
P_J^{\bf 4}&=&v^{J}(\lambda_1+\lambda_2+\lambda_3+\lambda_4)+
\frac{v^{J-1}}{2i}\[(\lambda_1+\lambda_2+\lambda_3+\lambda_4)(M-\bar M)
+\lambda _4 \left(\Delta
-S_1+S_2\right)
\]\\
\nn&+&\frac{v^{J-1}}{2i}\[\lambda _3 \left(\Delta
+S_1-S_2\right)+\lambda _2 \left(-\Delta
-S_1-S_2\right)+\lambda _1 \left(-\Delta
+S_1+S_2\right)\]+\dots\\
\la{Pasm} P_{2J}^{\bf 6}&=&\[v^{2J}+
\frac{v^{2J-1}}{i}(M-\bar M)\](\lambda_1\lambda_2+\lambda_1\lambda_3+\lambda_1\lambda_4+\lambda_2\lambda_3+\lambda_2\lambda_4+\lambda_3\lambda_4)\\
\nn&-&\frac{v^{2J-1}}{i}\[
\Delta\left(\lambda _1
   \lambda _2-\lambda _3 \lambda _4\right)+
S_2\left(\lambda _2 \lambda _3-\lambda _1
   \lambda _4\right)
   +S_1\left(\lambda _2 \lambda _4-\lambda
   _1 \lambda _3\right) \]+\dots\\
P_J^{\bar{\bf 4}}&=&v^{J}(\bar\lambda_1+\bar\lambda_2+\bar\lambda_3+\bar\lambda_4)+
\frac{v^{J-1}}{2i}\[(\bar\lambda_1+\bar\lambda_2+\bar\lambda_3+\bar\lambda_4)(M-\bar M)
-\bar\lambda _4 \left(\Delta
-S_1+S_2\right)\nn
\]
\\
\nn&-&\frac{v^{J-1}}{2i}\[\bar\lambda _3 \left(\Delta
+S_1-S_2\right)+\bar\lambda _2 \left(-\Delta
-S_1-S_2\right)+\bar\lambda _1 \left(-\Delta
+S_1+S_2\right)\]+\dots
\eeqa
where $\bar\lambda_a\equiv\frac{1}{\lambda_a}$.
We remind that $\lambda_i$'s are the diagonal 
elements of the twist matrix $G$.
Next, we plug the asymptotics
\eq{Pasm} into the TQ-relation
\eq{BaxterGeneral0} to deduce the following four 
possible asymptotics of the four linearly independent solutions of \eq{BaxterGeneral0}
\beq\la{qa}
q_a=\lambda_a^{-i v} v^{+\mu_a-\frac{J+M+\bar M}{2}}+\dots\ ,\qquad \mu_a=\frac{1}{2}\left(\bea{c}
+\Delta-S_1-S_2\\
+\Delta+S_1+S_2\\
-\Delta-S_1+S_2\\
-\Delta+S_1-S_2
\eea\right)_a\;.
\eeq

Another important asymptotic is the value  of $P_{2J}^{\bf 6}(v)$ at
the origin.
Due to the relation \eq{PinT}
and the observation that ${\mathbb T}^{\bf 6}(0)=H_q+1$ we have the following relation
\beq\la{atzero} 
P_{2J}^{\bf 6}(0)=\xi^{2J}\;.
\eeq
This is central for how the fishchain condition $H_q=0$ 
injects the information about the coupling $\xi$
into the TQ-relations. Except for \eq{atzero} there is no other explicit dependence on $\xi$ appearing in \eq{BaxterGeneral0}. 

\subsection{Discrete reparametrization symmetry and integrability}

The general expectation from the discrete reparametrization symmetry,
described in section~\ref{sec5}, 
is that the spectrum and other integrals of motion should remain invariant.
In this section we will prove this is indeed the case based on the integrability construction described in the previous section.

\paragraph{Magnon reordering symmetry.}
One can immediately notice that the Baxter equation \eq{BaxterGeneral0}
does not contain any explicit reference to the ordering of the particles. The equation only depends on the number of magnons $M$ and anti-magnons $\bar M$, but not on their particular ordering along the fishchain.
Furthermore, we will see below that the  Baxter TQ-relations depends non-trivially only on the difference $J_2=M-\bar M$.

At the level of the eigenvector this symmetry is less obvious. In section \ref{sec5} we have shown that wave functions with different magnon positions corresponds to different ways of cutting the same Feynman diagrams. Using this picture, we have argued that a shift in the magnon position should be thought of as a discrete reparametrization symmetry of the fishchain model. This shift is generated by the fishchain operator ${\mathbb H}_{3,2}$ in (\ref{Rop})
\beq
{\cal R}_{21}\equiv{\mathbb H}_{3,2}(Z_1,Z_2)=\nabla_1.\nabla_2+ \nabla_2.Z_1+2 Z_2.\nabla_1+2 Z_1.Z_2\;.
\eeq
We will now prove that indeed, this operator generates a symmetry of the transfer matrix. 
The operator ${\cal R}_{21}$ has the following key property
\beqa\la{RLL0}
&&{\mathbb L}^{ab}_2(u)
{\mathbb L}^{bc}_1\(u-\tfrac{i}{2\xi}\)
{\cal R}_{21}
-
{\cal R}_{21}
{\mathbb L}^{ab}_2\(u-\tfrac{i}{2\xi}\)
{\mathbb L}^{bc}_1(u)\\
\nn&&=\frac{i \Sigma^{ac}_{NM}}{2\xi^2}\[
( Z_2^N\nabla_1^M+ Z_2^N Z_1^M)\(\frac{\xi^2\;\tr\! q_2^2}{2}-4\)-
( Z_1^N\nabla_2^M+2 Z_1^N Z_2^M)\(\frac{\xi^2\;\tr\! q_1^2}{2}-3\)
\]\;.
\eeqa
At first sight it looks quite complicated, however,
one can see easily that on 
the states with a magnon on site $1$
and no magnon on $2$ the r.h.s. become zero and the
operator ${\cal R}$ become the interchange operator which exchanges the roles between the magnons and no-magnons.
The existence of such operators guarantees that the spectrum of all integrals of motion does not depend on the initial order of the magnons. This ensures the self-consistency of our approach.

More precisely we can see that
the magnon move operation results in the following similarity transformation
\beq\la{simi}
\hat {\mathbb T}^{\bf I}(u)\,|\Psi\>^{\bf I}=
({\cal R}_{k,k+1})^{-1}\,\hat {\mathbb T}^{\overset\leftrightarrow{\bf I}}(u)
\,{\cal R}_{k,k+1}\,|\Psi\>^{\bf I}=
({\cal R}_{k,k+1})^{-1}\,\hat {\mathbb T}^{\overset\leftrightarrow{\bf I}}(u)\,|\Psi\>^{\overset\leftrightarrow{\bf I}}
\eeq
where the set $\bf I$ is such that $I_k=1$
and $I_{k+1}=0$, whereas the transformed set
$\overset\leftrightarrow{\bf I}$
corresponds to the magnon at $k+1$ i.e.
$I_k=0$
and $I_{k+1}=1$.
The equation \eq{simi} ensures that the spectrum of all conserved charges
before and after the magnon move operation stays unchanged. Similarly, we can shift an anti-magnon to the left and all conserved charges are unaffected. Since states are uniquely characterize by the charges, this is a gauge symmetry that is associated with a redundancy in our description.

\paragraph{Magnon--anti-magnon annihilation.}
First, let us demonstrate that the spectrum 
does not change when we simultaneously remove one magnon and one anti-magnon.
We will establish a simple relation between the solutions of the Baxter equation with different numbers of magnons.
Consider the transformation $M\to M-n,\;\bar M\to \bar M-n$. Under this transformation \eq{BaxterGeneral0} becomes
\beqa\la{BaxterGeneral2}
-\frac{P_{2J}^{\bf 6}(v)}{v^{J-M-\bar M+2n}}\;\tilde q_a&=&\frac{v^M (v+i)^J (v+2i)^{\bar M}}{v^n(v+2i)^n}\; \tilde q_a^{[+4]}
-
\frac{v^M(v+i)^{\bar M}}{v^n(v+i)^{n}}P_J^{\bf 4}(v+\tfrac i2)\;\tilde q_a^{[+2]}\\
\nn
&+&
\frac{v^{\bar M}(v-i)^{J}(v-2i)^M}{v^n(v-2i)^n}\;\tilde q_a^{[-4]}
-
\frac{v^{\bar M}(v-i)^M P_J^{\bar{\bf 4}}(v-\tfrac i2)}{v^n(v-i)^n}\;\tilde q_a^{[-2]}
\;.
\eeqa
First we see that the overall factor $1/v^n$ cancels. Furthermore, by denoting
$q(v)=\frac{\tilde q(v)}{v^n}$ we return back to the initial form
\eq{BaxterGeneral0}. 

At the level of the fishchain wave function, we can always shift the positions of the magnon and anti-magnon until the anti-magnon is at the first site and the magnon is at the second site. To Annihilate them, we act with
\beq
{\mathbb H}_{2,2}(Z_1,Z_2)\propto \tilde{\cal R}_{12}= \nabla_{1}.\nabla_{2}+ 2Z_2.\nabla_{1}+2Z_1.\nabla_{2}+4 Z_1.Z_2\;.
\eeq
This operator has the following property
\beqa
&&{\mathbb L}^{ab}_1(u)
{\mathbb L}^{bc}_2\(u\)
\tilde{\cal R}_{12}
-
\tilde{\cal R}_{12}
{\mathbb L}^{ab}_1(u-\tfrac{i}{2\xi})
{\mathbb L}^{bc}_2(u+\tfrac{i}{2\xi})\\
\nn&&=\frac{i \Sigma^{ac}_{NM}}{2\xi^2}\[
( Z_1^N{\cal D}_2^M+ 2Z_1^N Z_2^M)\(\frac{\xi^2\;\tr\! q_1^2}{2}-4\)-
( Z_2^N{\cal D}_1^M+2 Z_2^N Z_1^M)\(\frac{\xi^2\;\tr\! q_2^2}{2}-4\)
\]\;.
\eeqa
Similar to \eq{RLL0}
the r.h.s. of the above equation vanishes on the states with 
magnon and anti-magnon. This relation leads to an analog of
\eq{simi} where the initial set ${\bf I}$
has $I_{k+1}=1$ and $I_k=-1$, whereas the resulting set
${\overset\leftrightarrow{\bf I}}$ has no magnons at these positions i.e.
${\overset\leftrightarrow{ I}}_k={\overset\leftrightarrow{ I}}_{k+1}=0$. We see again that the magnon annihilation operation
does not affect the spectrum or any of the the conserved charges.

\subsection{Quantization condition and exact spectrum}\la{secnm}
In order to adapt the quantization condition of
\cite{Gromov:2019cja} to the general case with any number of magnons, we introduce two sets of solutions of the TQ-relations \eq{BaxterGeneral0}. 
One set, denoted by $q_a^{\downarrow}(v)$, is regular in the upper half-plane and satisfy \eq{qa} at large $v$. The other set, $q_a^{\uparrow}(v)$, is regular in the lower half-plane. It is easy to see that such solutions exist and that they can be constructed numerically, using the methods of \cite{Gromov:2015wca}.

Since there are only $4$ independent solutions of the $4$th order finite difference equation \eq{BaxterGeneral0}, 
these two sets $q_a^{\downarrow}(v)$
and $q_a^{\uparrow}(v)$ should be linearly dependent with i-periodic coefficients, i.e.
\beq\la{OmeDef}
q_a^{\uparrow}(v) = {\Omega_a}^b(v)\; q_b^{\downarrow}(v)\ ,\qquad{\Omega_a}^b(v+i)={\Omega_a}^b(v) \;.
\eeq
The matrix ${\Omega_a}^b(v)$ itself can be constructed from q-functions as the simple ratio of two determinants
\beq
{\Omega_a}^b(v)=
\frac{\epsilon^{b\;b_1 b_2 b_3}}{3!}\frac{
\det\limits_{n=0,\dots,3}\{q_a^\uparrow(v-i n), q_{b_1}^\downarrow(v-i n),q_{b_2}^\downarrow(v-i n),q_{b_3}^\downarrow(v-i n)\}
}{
\det\limits_{n=0,\dots,3}\{q_{1}^\downarrow(v-i n), q_{2}^\downarrow(v-i n),q_{3}^\downarrow(v-i n),q_{4}^\downarrow(v-i n)\}
}\;,
\eeq
as follows from \eq{OmeDef}. 

In \cite{Gromov:2019cja} it is shown that this matrix ${\Omega_a}^b(v)$ should satisfy the following {\it quantization condition}\footnote{We are grateful to the authors of \cite{Gromov:2019cja} for sharing this unpublished result with us}
\beq\la{Omegacond}
{\Omega_1}^2={\Omega_2}^1={\Omega_3}^4
={\Omega_4}^3=0\;,
\eeq
which singles out physical solutions of the TQ-relations.
Furthermore, one can show \cite{Gromov:2019cja}
that the $i$-periodic function ${\Omega_a}^b(u)$ has simple poles at $u= i n$ of order $J$. This implies that
each matrix element of 
${\Omega_a}^b(u)$ can be parametrized by $J+1$ 
constants $C^{(n)}$ as
\beq
{\Omega_a}^b(u)=\frac{\sum\limits_{n=0}^J {C^{(n)\;\;b}_{\;\;\;\;\;a}}\; e^{2\pi n u}}{(1-e^{2\pi u})^J}\;,
\eeq
where in addition for off-diagonal elements only $J$ constants $C^{(n)}$ are non-zero. This implies that we have $4J$ equations in \eq{Omegacond}.
One can show that only $4J-3$ of these conditions
are linearly independent (for large enough $J$), which is exactly how many free parameters we have in the polynomial coefficients $P_L$ in \eq{BaxterGeneral0}. Thus, using the quantization condition \eq{Omegacond} we should be able to fix all coefficients in the Baxter TQ-relations, which also includes $\Delta$.

In the next section we report on the numerical tests of this procedure for the case with magnons.

\subsection{Numerical tests}
In this section we apply the procedure described in section~\ref{secnm} for two cases, $J=3,\;M=1$ and $J=3,\;M=2$. In order to solve the Baxter TQ-relations \eq{BaxterGeneral0} we use the method of \cite{Gromov:2015wca,Gromov:2017cja}.
Then we use the generalized gluing conditions \eq{Omegacond} to fix the parameters in \eq{BaxterGeneral0}.
In both cases we found perfect agreement with the Asymptotic Bethe Ansatz (ABA) of \cite{Caetano:2016ydc,Ipsen:2018fmu,Beisert:2006ez} at weak coupling. In this section we fix the twist parameters as
$\lambda_1=e^{i/3},\;\lambda_2=e^{-i/3},\;\lambda_3=e^{i/2},\;\lambda_4=e^{-i/2}$.

\subsubsection{Length three, one magnon}
\begin{figure}[h]
    \centering
    \includegraphics[scale=0.75]{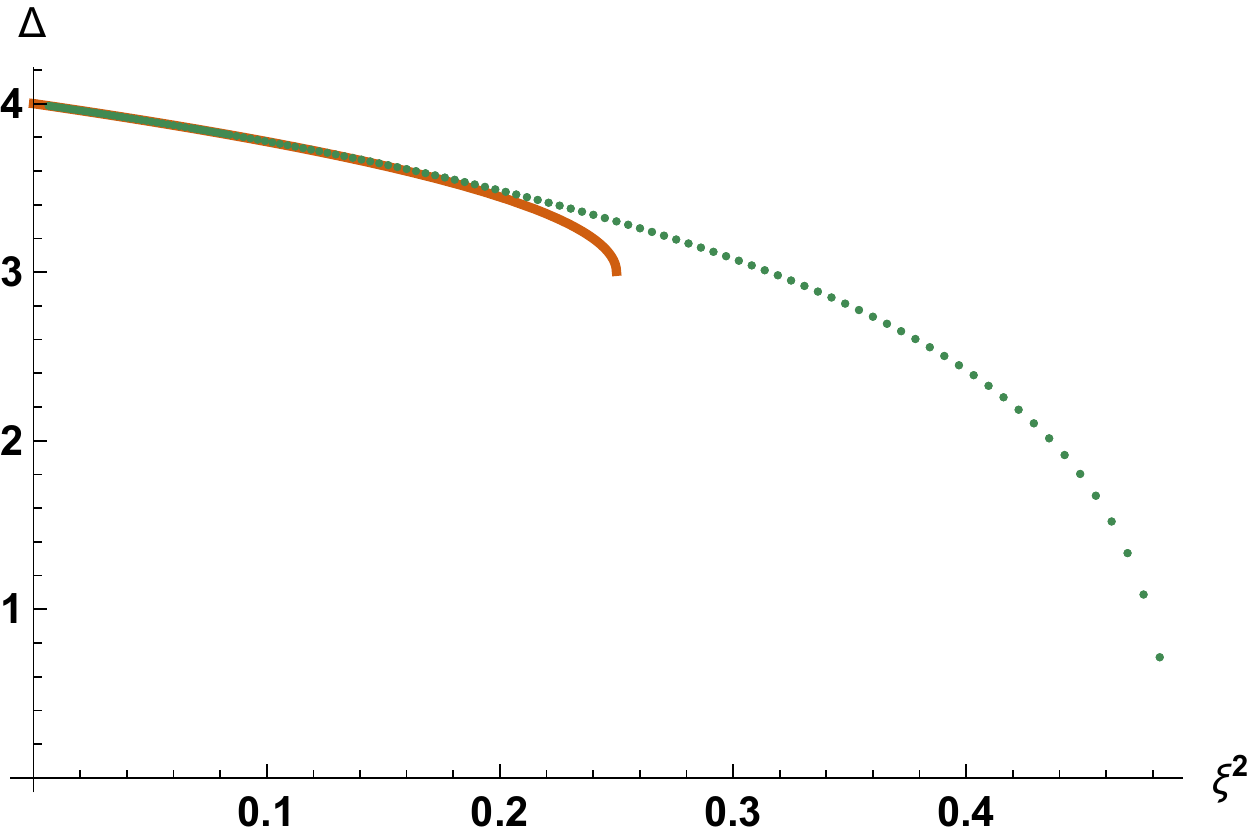}
    \caption{Numerical solution (dots) and the Asymptotic Bethe Ansatz prediction (solid line) match perfectly at small coupling for the state with one magnon and $J=3$. The two start to deviate from each other at larger values of $\xi^2$.}
    \label{fig:J3M1}
\end{figure}
We generated to a very high precision numerical values for a wide range of couplings using the method described in~\ref{secnm} (see figure~\ref{fig:J3M1}).
In order to check our results we have to recall
that in the ABA approach
for one magnon one only needs to know the
{\it dispersion relation}
$\Delta_{ABA}=J+\sqrt{1-4 \xi ^2}$,
which is expected to be accurate up to the wrapping order $\xi^{2J+2M}=\xi^8$. 

Indeed, by making a fit of our numerical values with even
powers of $\xi$ we found the following expansion
\beqa
\Delta_{TQ}=4.-2. \xi ^2-2. \xi ^4-4. \xi ^6+0.155845 \xi ^8+\dots\;.
\eeqa
We see that the first $4$ coefficients are integers, and agree precisely with the magnon dispersion relation. At order $\xi^8$ we get some
real number, which is expected to be a combination of two polylogarithms ${\rm Li}_3$ and ${\rm Li}_5$ with the arguments depending on the twists.

\begin{figure}[t]
    \centering
    \includegraphics[scale=0.3]{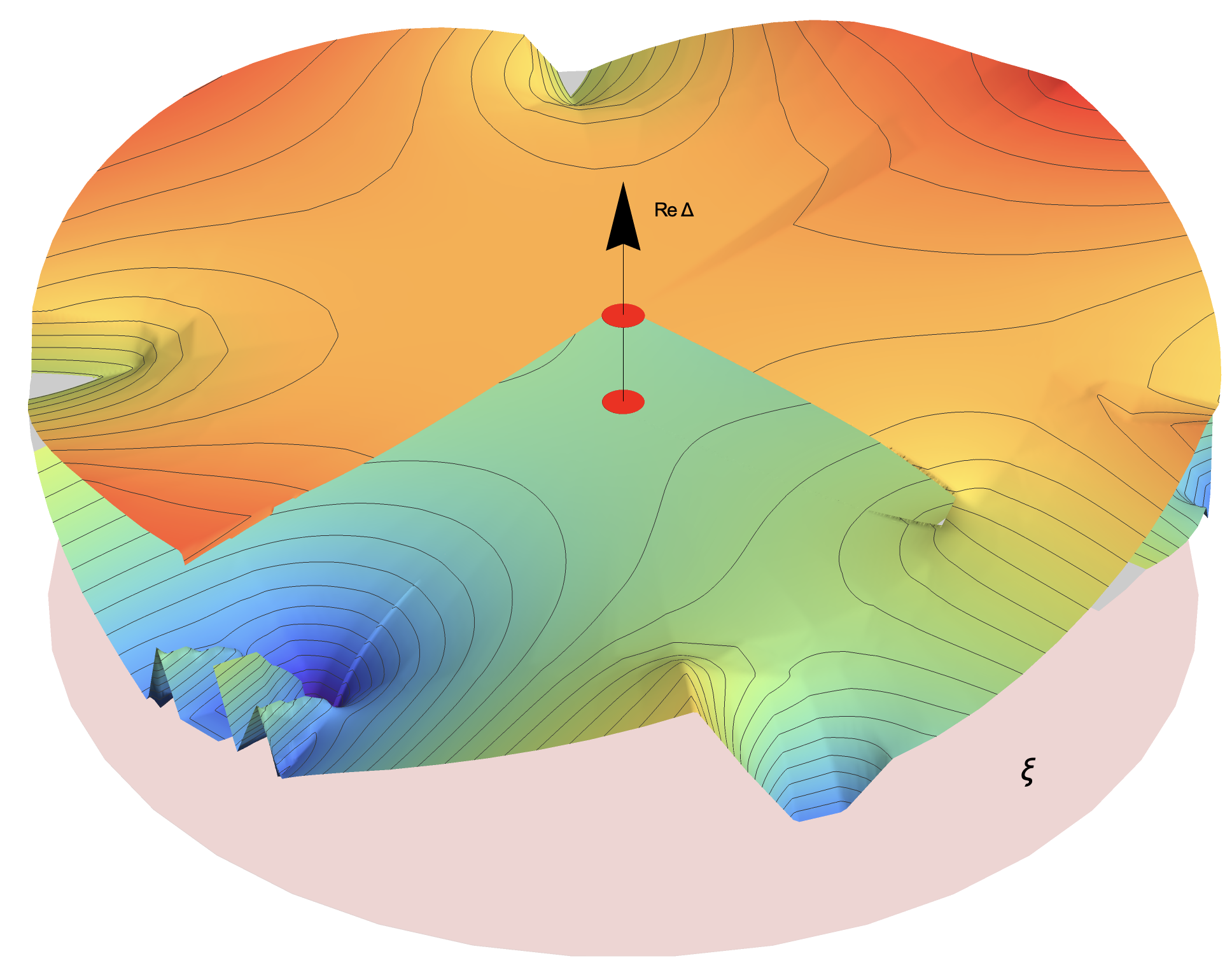}
    \caption{Single magnon operator $\tr \phi_1^3\phi_2$  has a dimension $\Delta$, which is 
 a multivalued function of $\xi$. At weak coupling
 the bare dimension of the state is $J_1+J_2=4$. Analytically continuing in $\xi$ one can discover an infinite tower of states with the bare dimensions $4+2n,\;n=0,1,\dots$ ($n=0,1$ are indicated by the red dots on the picture).}
    \label{fig:mag}
\end{figure}
What is very interesting is the analytic structure of our result in $\xi$.
At small enough $\xi$ the function has a regular (convergent) expansion in powers of $\xi^2$.
At $\xi\sim 0.7$ the dimension $\Delta$
tends to $0$ and after that point become purely imaginary (see figure~\ref{fig:J3M1}). The value $\xi\sim 0.7$ is thus a branch point. This is not the only branch cut of the function $\Delta(\xi)$. The function $\Delta(\xi)$ is in fact a multivalued function with infinitely many sheets (see figure~\ref{fig:mag}), containing not just one but infinitely many local operators, which can be identified at weak coupling as $\tr{\phi_1^3\phi_2(\phi_2\phi_2^\dagger)^n}$ (up to permutations).

\subsubsection{Length  three, two magnons}
We also found a numerical solution in the two magnon case for the chain of length $J=3$.
Again our results are in complete agreement with the ABA prediction at weak coupling.
Note that this time, the ABA calculation involves a rather non-trivial dressing phase, which gives rise to the $\zeta$ values in the result
\beqa\la{J3aba}
\Delta_{ABA}&=&5+\left(\sqrt{5}-1\right) \xi ^2
+\left(\frac{3}{\sqrt{5}}-1\right) \xi ^4
-\left(2+\frac{2}{5 \sqrt{5}}\right) \xi^6
\\ \nn
&+&\left(-\frac{4 \zeta _3}{\sqrt{5}}-4 \zeta _3+\frac{81}{25 \sqrt{5}}+3\right) \xi ^8
+\left(\frac{112 \zeta _3}{5
   \sqrt{5}}-\frac{424}{25 \sqrt{5}}+8\right) \xi ^{10}+{\cal O}\left(\xi ^{12}\right)\;.
\eeqa
Our numerical fit of the result of 
solution of the TQ-relations gives
\beqa\la{J3num}
\Delta_{TQ}=
5+1.23607 \xi ^2+0.341641 \xi ^4-2.17889 \xi ^6-2.50956 \xi ^8-0.575805 \xi ^{10}+\dots
\eeqa
which agrees perfectly with the first $4$-loops predicted by ABA \eq{J3aba}. At 
the wrapping order $\xi^{2J+2M}=\xi^{10}$, as expected,
the ABA result is no longer valid and deviates from our numerical prediction \eq{J3num}.

In conclusion, the fact that we managed to reproduce the ABA prediction from the TQ-relations provides a very non-trivial test of our approach.

\section{Conclusion}\la{sec7}

In this paper, we completed the derivation
of the Holographic dual of the fishnet theory.
Introducing the magnons, i.e. $\phi_2$ and/or $\phi_2^\dagger$ fields, into the consideration brought some surprises.
Firstly, we found that the classical limit $\xi\to\infty$ remain unchanged. Namely, we still
find the same classically integrable fishchain model, which is a classical chain of particles on a light-cone with nearest-neighbor interaction. 
The essence of the magnons is hiding in the details of the quantization procedure, which also modifies the light-cone into the $AdS_5$
space.\footnote{$AdS_5$ space first appeared in the regime of large R-charge of the fishnet model in \cite{Basso:2018agi}.} This is, in fact, in complete agreement with the exact spectrum and 4-point function computation in~\cite{Gromov:2018hut} for the length-two operators.
Secondly, the introduction of the magnons revealed a new nice feature, namely the {\it discrete reparametrization symmetry} -- an invariance of the physics w.r.t. different sectioning of the discretized world-sheet of the fishchain into space-like slices. The magnons in this picture indicate bends of the curve cutting the world-sheet. Mathematically this symmetry manifests as a map between Hilbert spaces,
preserving the integrability and conserved charges that uniquely characterize the state. 

Another important output of our analysis is the universal integrability picture, which includes all possible insertions on equal footing. We found the quantum Baxter TQ-relations, which together with the quantization condition~\cite{Gromov:2017cja,Gromov:2019cja}, give us direct access to the quantum spectrum of the model in complete generality.\footnote{This spectrum was previously only available for a rather narrow subset of states~\cite{Gromov:2017cja}.} This also opens ways to the separation of variables (SoV) construction, along the lines of the recent
works~\cite{Gromov:2016itr,Ryan:2018fyo,Derkachov:2018rot,Cavaglia:2019pow}.

Among open questions remains a particular configuration when the number of magnons saturates the number of states, i.e. when both $\mathfrak{u}(1)$ charges are equal $|J_1|=|J_2|$. In this case, surprisingly, the general strategy outlined in this paper fails.
On some particular examples it is known that this case produces a more complicated result for the spectrum even for short operators, tractable with the conformal symmetry alone~\cite{Gromov:2018hut}. Nevertheless, the integrability should be present in this case too~\cite{Gromov:2019cja}, but the strong coupling behavior is very different, indicating that it is mysteriously missing in the current picture. This problem could be related to another large part of states which happened to be protected, and decoupled from the dynamical part of the spectrum studied in this paper. The reason why there is a large number of states with protected dimension is also unclear, there is no manifest symmetry, like supersymmetry, which can help to explain this phenomenon.

\section*{Acknowledgements}

We thank O.~Aharony, D.~Grabner, V.~Kazakov, G.~Korchemsky and A.~Zhiboedov for invaluable discussions. We thank J.~McGovern for comments on the manuscript. N.G. was supported by the STFC grant (ST/P000258/1). A.S. was supported by the I-CORE Program of the Planning and Budgeting Committee, The Israel Science Foundation (1937/12).

\end{document}

%% file: graphs_building_magnons4.pdf_tex
\begingroup%
  \makeatletter%
  \providecommand\color[2][]{%
    \errmessage{(Inkscape) Color is used for the text in Inkscape, but the package 'color.sty' is not loaded}%
    \renewcommand\color[2][]{}%
  }%
  \providecommand\transparent[1]{%
    \errmessage{(Inkscape) Transparency is used (non-zero) for the text in Inkscape, but the package 'transparent.sty' is not loaded}%
    \renewcommand\transparent[1]{}%
  }%
  \providecommand\rotatebox[2]{#2}%
  \ifx\svgwidth\undefined%
    \setlength{\unitlength}{316.83229461bp}%
    \ifx\svgscale\undefined%
      \relax%
    \else%
      \setlength{\unitlength}{\unitlength * \real{\svgscale}}%
    \fi%
  \else%
    \setlength{\unitlength}{\svgwidth}%
  \fi%
  \global\let\svgwidth\undefined%
  \global\let\svgscale\undefined%
  \makeatother%
  \begin{picture}(1,0.46695538)%
    \put(0,0){\includegraphics[width=\unitlength,page=1]{graphs_building_magnons4.pdf}}%
    \put(0.12952269,0.45069016){\color[rgb]{0,0,0}\makebox(0,0)[lb]{\smash{$x_J$}}}%
    \put(0.23052254,0.45069016){\color[rgb]{0,0,0}\makebox(0,0)[lb]{\smash{$x_i$}}}%
    \put(0.33152234,0.45069016){\color[rgb]{0,0,0}\makebox(0,0)[lb]{\smash{$x_{i-1}$}}}%
    \put(0.43252215,0.45069016){\color[rgb]{0,0,0}\makebox(0,0)[lb]{\smash{$x_{i-2}$}}}%
    \put(0.53352195,0.45069016){\color[rgb]{0,0,0}\makebox(0,0)[lb]{\smash{$x_k$}}}%
    \put(0.63452176,0.45069016){\color[rgb]{0,0,0}\makebox(0,0)[lb]{\smash{$x_{k-1}$}}}%
    \put(0.73552156,0.45069016){\color[rgb]{0,0,0}\makebox(0,0)[lb]{\smash{$x_{k-2}$}}}%
    \put(0.83652137,0.45069016){\color[rgb]{0,0,0}\makebox(0,0)[lb]{\smash{$x_2$}}}%
    \put(0.93752117,0.45069016){\color[rgb]{0,0,0}\makebox(0,0)[lb]{\smash{$x_1$}}}%
    \put(0.12952269,0.20829062){\color[rgb]{0,0,0}\makebox(0,0)[lb]{\smash{$y_J$}}}%
    \put(0.23052254,0.20829062){\color[rgb]{0,0,0}\makebox(0,0)[lb]{\smash{$y_i$}}}%
    \put(0.33152234,0.20829062){\color[rgb]{0,0,0}\makebox(0,0)[lb]{\smash{$y_{i-1}$}}}%
    \put(0.43252215,0.20829062){\color[rgb]{0,0,0}\makebox(0,0)[lb]{\smash{$y_{i-2}$}}}%
    \put(0.53352195,0.20829062){\color[rgb]{0,0,0}\makebox(0,0)[lb]{\smash{$y_k$}}}%
    \put(0.63452176,0.20829062){\color[rgb]{0,0,0}\makebox(0,0)[lb]{\smash{$y_{k-1}$}}}%
    \put(0.73552156,0.20829062){\color[rgb]{0,0,0}\makebox(0,0)[lb]{\smash{$y_{k-2}$}}}%
    \put(0.83652137,0.20829062){\color[rgb]{0,0,0}\makebox(0,0)[lb]{\smash{$y_2$}}}%
    \put(0.93752117,0.20829062){\color[rgb]{0,0,0}\makebox(0,0)[lb]{\smash{$y_1$}}}%
    \put(-0.00177705,0.31434042){\color[rgb]{0,0,0}\makebox(0,0)[lb]{\smash{$\hat\cB=$}}}%
  \end{picture}%
\endgroup%

%% file: magnon_move2.pdf_tex
\begingroup%
  \makeatletter%
  \providecommand\color[2][]{%
    \errmessage{(Inkscape) Color is used for the text in Inkscape, but the package 'color.sty' is not loaded}%
    \renewcommand\color[2][]{}%
  }%
  \providecommand\transparent[1]{%
    \errmessage{(Inkscape) Transparency is used (non-zero) for the text in Inkscape, but the package 'transparent.sty' is not loaded}%
    \renewcommand\transparent[1]{}%
  }%
  \providecommand\rotatebox[2]{#2}%
  \ifx\svgwidth\undefined%
    \setlength{\unitlength}{630.2865961bp}%
    \ifx\svgscale\undefined%
      \relax%
    \else%
      \setlength{\unitlength}{\unitlength * \real{\svgscale}}%
    \fi%
  \else%
    \setlength{\unitlength}{\svgwidth}%
  \fi%
  \global\let\svgwidth\undefined%
  \global\let\svgscale\undefined%
  \makeatother%
  \begin{picture}(1,0.49385887)%
    \put(0,0){\includegraphics[width=\unitlength,page=1]{magnon_move2.pdf}}%
    \put(0.48257379,0.30151573){\color[rgb]{0,0,0}\makebox(0,0)[lb]{\smash{$k-1$}}}%
    \put(-0.000814,0.41073569){\color[rgb]{0,0,0}\makebox(0,0)[lb]{\smash{$(x_{k+1,k}^2\Box_{k+1})\,(\Box_kx_{k,k-1}^2)\,\Box_{k-1}$}}}%
    \put(0.44195735,0.30151573){\color[rgb]{0,0,0}\makebox(0,0)[lb]{\smash{$k$}}}%
    \put(0.37087857,0.30151573){\color[rgb]{0,0,0}\makebox(0,0)[lb]{\smash{$k+1$}}}%
    \put(0.71182348,0.55239461){\color[rgb]{0,0,0}\makebox(0,0)[lt]{\begin{minipage}{1.04848349\unitlength}\raggedright \end{minipage}}}%
    \put(0,0){\includegraphics[width=\unitlength,page=2]{magnon_move2.pdf}}%
    \put(0.04487953,0.11118945){\color[rgb]{0,0,0}\makebox(0,0)[lb]{\smash{$(x_{k+1,k}^2\Box_{k+1})\,(\Box_kx_{k,k-1}^2)$}}}%
    \put(0,0){\includegraphics[width=\unitlength,page=3]{magnon_move2.pdf}}%
    \put(0.48257379,0.0019695){\color[rgb]{0,0,0}\makebox(0,0)[lb]{\smash{$k-1$}}}%
    \put(0.44195735,0.0019695){\color[rgb]{0,0,0}\makebox(0,0)[lb]{\smash{$k$}}}%
    \put(0.37087859,0.0019695){\color[rgb]{0,0,0}\makebox(0,0)[lb]{\smash{$k+1$}}}%
    \put(0.93950878,0.30151577){\color[rgb]{0,0,0}\makebox(0,0)[lb]{\smash{$k-1$}}}%
    \put(0.89889234,0.30151577){\color[rgb]{0,0,0}\makebox(0,0)[lb]{\smash{$k$}}}%
    \put(0.82781356,0.30151577){\color[rgb]{0,0,0}\makebox(0,0)[lb]{\smash{$k+1$}}}%
    \put(0,0){\includegraphics[width=\unitlength,page=4]{magnon_move2.pdf}}%
    \put(0.64651059,0.11118945){\color[rgb]{0,0,0}\makebox(0,0)[lb]{\smash{$(x_{k+1,k}^2\Box_{k+1})$}}}%
    \put(0.93443172,0.0019695){\color[rgb]{0,0,0}\makebox(0,0)[lb]{\smash{$k-1$}}}%
    \put(0.89381528,0.0019695){\color[rgb]{0,0,0}\makebox(0,0)[lb]{\smash{$k$}}}%
    \put(0.8227365,0.0019695){\color[rgb]{0,0,0}\makebox(0,0)[lb]{\smash{$k+1$}}}%
    \put(0,0){\includegraphics[width=\unitlength,page=5]{magnon_move2.pdf}}%
  \end{picture}%
\endgroup%

%% file: cut_deformation3.pdf_tex
\begingroup%
  \makeatletter%
  \providecommand\color[2][]{%
    \errmessage{(Inkscape) Color is used for the text in Inkscape, but the package 'color.sty' is not loaded}%
    \renewcommand\color[2][]{}%
  }%
  \providecommand\transparent[1]{%
    \errmessage{(Inkscape) Transparency is used (non-zero) for the text in Inkscape, but the package 'transparent.sty' is not loaded}%
    \renewcommand\transparent[1]{}%
  }%
  \providecommand\rotatebox[2]{#2}%
  \ifx\svgwidth\undefined%
    \setlength{\unitlength}{572.78726893bp}%
    \ifx\svgscale\undefined%
      \relax%
    \else%
      \setlength{\unitlength}{\unitlength * \real{\svgscale}}%
    \fi%
  \else%
    \setlength{\unitlength}{\svgwidth}%
  \fi%
  \global\let\svgwidth\undefined%
  \global\let\svgscale\undefined%
  \makeatother%
  \begin{picture}(1,0.25292455)%
    \put(0,0){\includegraphics[width=\unitlength,page=1]{cut_deformation3.pdf}}%
    \put(0.09380572,0.24662745){\color[rgb]{0,0,0}\makebox(0,0)[lb]{\smash{$\chi_0$}}}%
    \put(0.14967288,0.24662745){\color[rgb]{0,0,0}\makebox(0,0)[lb]{\smash{$\chi_0$}}}%
    \put(0.20554004,0.24662745){\color[rgb]{0,0,0}\makebox(0,0)[lb]{\smash{$\chi_0$}}}%
    \put(0.26140721,0.24662745){\color[rgb]{0,0,0}\makebox(0,0)[lb]{\smash{$\chi_0$}}}%
    \put(0.31727437,0.24662745){\color[rgb]{0,0,0}\makebox(0,0)[lb]{\smash{$\chi_0$}}}%
    \put(0.37314153,0.24662745){\color[rgb]{0,0,0}\makebox(0,0)[lb]{\smash{$\chi_0$}}}%
    \put(0.65247731,0.24662745){\color[rgb]{0,0,0}\makebox(0,0)[lb]{\smash{$\chi_0$}}}%
    \put(0.70834447,0.24662745){\color[rgb]{0,0,0}\makebox(0,0)[lb]{\smash{$\chi_0$}}}%
    \put(0.7530382,0.24662745){\color[rgb]{0,0,0}\makebox(0,0)[lb]{\smash{$\chi_{-1}$}}}%
    \put(0.8200788,0.24662745){\color[rgb]{0,0,0}\makebox(0,0)[lb]{\smash{$\chi_1$}}}%
    \put(0.87594596,0.24662745){\color[rgb]{0,0,0}\makebox(0,0)[lb]{\smash{$\chi_0$}}}%
    \put(0.93181313,0.24662745){\color[rgb]{0,0,0}\makebox(0,0)[lb]{\smash{$\chi_0$}}}%
    \put(0,0){\includegraphics[width=\unitlength,page=2]{cut_deformation3.pdf}}%
  \end{picture}%
\endgroup%

%% file: vact.pdf_tex
\begingroup%
  \makeatletter%
  \providecommand\color[2][]{%
    \errmessage{(Inkscape) Color is used for the text in Inkscape, but the package 'color.sty' is not loaded}%
    \renewcommand\color[2][]{}%
  }%
  \providecommand\transparent[1]{%
    \errmessage{(Inkscape) Transparency is used (non-zero) for the text in Inkscape, but the package 'transparent.sty' is not loaded}%
    \renewcommand\transparent[1]{}%
  }%
  \providecommand\rotatebox[2]{#2}%
  \ifx\svgwidth\undefined%
    \setlength{\unitlength}{771.8640329bp}%
    \ifx\svgscale\undefined%
      \relax%
    \else%
      \setlength{\unitlength}{\unitlength * \real{\svgscale}}%
    \fi%
  \else%
    \setlength{\unitlength}{\svgwidth}%
  \fi%
  \global\let\svgwidth\undefined%
  \global\let\svgscale\undefined%
  \makeatother%
  \begin{picture}(1,0.42338333)%
    \put(0,0){\includegraphics[width=\unitlength,page=1]{vact.pdf}}%
    \put(0.11816381,0.24274232){\color[rgb]{0,0,0}\makebox(0,0)[lb]{\smash{$k$}}}%
    \put(0.11922345,0.41535541){\color[rgb]{0,0,0}\makebox(0,0)[lb]{\smash{$v_k^{-1}$}}}%
    \put(0,0){\includegraphics[width=\unitlength,page=2]{vact.pdf}}%
    \put(0.86440923,0.24274232){\color[rgb]{0,0,0}\makebox(0,0)[lb]{\smash{$k$}}}%
    \put(0,0){\includegraphics[width=\unitlength,page=3]{vact.pdf}}%
    \put(0.49128646,0.24274232){\color[rgb]{0,0,0}\makebox(0,0)[lb]{\smash{$k$}}}%
    \put(0,0){\includegraphics[width=\unitlength,page=4]{vact.pdf}}%
    \put(0.11816381,0.00228546){\color[rgb]{0,0,0}\makebox(0,0)[lb]{\smash{$k$}}}%
    \put(0,0){\includegraphics[width=\unitlength,page=5]{vact.pdf}}%
    \put(0.11922345,0.17697146){\color[rgb]{0,0,0}\makebox(0,0)[lb]{\smash{$v_k$}}}%
    \put(0,0){\includegraphics[width=\unitlength,page=6]{vact.pdf}}%
    \put(0.86440923,0.00228546){\color[rgb]{0,0,0}\makebox(0,0)[lb]{\smash{$k$}}}%
    \put(0,0){\includegraphics[width=\unitlength,page=7]{vact.pdf}}%
    \put(0.49128652,0.00228546){\color[rgb]{0,0,0}\makebox(0,0)[lb]{\smash{$k$}}}%
    \put(0,0){\includegraphics[width=\unitlength,page=8]{vact.pdf}}%
  \end{picture}%
\endgroup%

%% file: hact.pdf_tex
\begingroup%
  \makeatletter%
  \providecommand\color[2][]{%
    \errmessage{(Inkscape) Color is used for the text in Inkscape, but the package 'color.sty' is not loaded}%
    \renewcommand\color[2][]{}%
  }%
  \providecommand\transparent[1]{%
    \errmessage{(Inkscape) Transparency is used (non-zero) for the text in Inkscape, but the package 'transparent.sty' is not loaded}%
    \renewcommand\transparent[1]{}%
  }%
  \providecommand\rotatebox[2]{#2}%
  \ifx\svgwidth\undefined%
    \setlength{\unitlength}{419.86393966bp}%
    \ifx\svgscale\undefined%
      \relax%
    \else%
      \setlength{\unitlength}{\unitlength * \real{\svgscale}}%
    \fi%
  \else%
    \setlength{\unitlength}{\svgwidth}%
  \fi%
  \global\let\svgwidth\undefined%
  \global\let\svgscale\undefined%
  \makeatother%
  \begin{picture}(1,0.32485365)%
    \put(0,0){\includegraphics[width=\unitlength,page=1]{hact.pdf}}%
    \put(0.82694979,0.00420143){\color[rgb]{0,0,0}\makebox(0,0)[lb]{\smash{$k$}}}%
    \put(0,0){\includegraphics[width=\unitlength,page=2]{hact.pdf}}%
    \put(0.21722846,0.00420143){\color[rgb]{0,0,0}\makebox(0,0)[lb]{\smash{$k$}}}%
    \put(0,0){\includegraphics[width=\unitlength,page=3]{hact.pdf}}%
    \put(0.18106882,0.31009538){\color[rgb]{0,0,0}\makebox(0,0)[lb]{\smash{$h_{k+1,k}^{-1}$}}}%
    \put(0,0){\includegraphics[width=\unitlength,page=4]{hact.pdf}}%
  \end{picture}%
\endgroup%

%% file: basic_moves6.pdf_tex
\begingroup%
  \makeatletter%
  \providecommand\color[2][]{%
    \errmessage{(Inkscape) Color is used for the text in Inkscape, but the package 'color.sty' is not loaded}%
    \renewcommand\color[2][]{}%
  }%
  \providecommand\transparent[1]{%
    \errmessage{(Inkscape) Transparency is used (non-zero) for the text in Inkscape, but the package 'transparent.sty' is not loaded}%
    \renewcommand\transparent[1]{}%
  }%
  \providecommand\rotatebox[2]{#2}%
  \ifx\svgwidth\undefined%
    \setlength{\unitlength}{771.86394639bp}%
    \ifx\svgscale\undefined%
      \relax%
    \else%
      \setlength{\unitlength}{\unitlength * \real{\svgscale}}%
    \fi%
  \else%
    \setlength{\unitlength}{\svgwidth}%
  \fi%
  \global\let\svgwidth\undefined%
  \global\let\svgscale\undefined%
  \makeatother%
  \begin{picture}(1,0.18707225)%
    \put(0,0){\includegraphics[width=\unitlength,page=1]{basic_moves6.pdf}}%
    \put(0.11898726,0.00228543){\color[rgb]{0,0,0}\makebox(0,0)[lb]{\smash{$k$}}}%
    \put(0.09931786,0.17904436){\color[rgb]{0,0,0}\makebox(0,0)[lb]{\smash{$r_{k+1,k}$}}}%
    \put(0,0){\includegraphics[width=\unitlength,page=2]{basic_moves6.pdf}}%
    \put(0.49211003,0.00228543){\color[rgb]{0,0,0}\makebox(0,0)[lb]{\smash{$k$}}}%
    \put(0,0){\includegraphics[width=\unitlength,page=3]{basic_moves6.pdf}}%
    \put(0.86523283,0.00228543){\color[rgb]{0,0,0}\makebox(0,0)[lb]{\smash{$k$}}}%
    \put(0,0){\includegraphics[width=\unitlength,page=4]{basic_moves6.pdf}}%
  \end{picture}%
\endgroup%